\documentclass[reprint, superscriptaddress, nobibnotes, amsmath, amssymb, aps, prb, floatfix]{revtex4-2}
\usepackage{xr}
\usepackage{hyperref}
\hypersetup{colorlinks=true, linkcolor=blue, filecolor=blue, urlcolor=blue, citecolor=blue}
\usepackage{cleveref}
\usepackage{multirow}
\usepackage{graphicx}
\usepackage{dcolumn}
\usepackage{bm}
\usepackage{xcolor}
\usepackage{textcomp}
\DeclareUnicodeCharacter{2212}{\ensuremath{-}}

\newcommand{\etal}{{\textit{et al. }}}

\usepackage{xcolor}

\begin{document}

\title{Revealing the Void-Size Distribution of Silica Glass using Persistent Homology}

\author{Achraf Atila}
\email{achraf.atila@bam.de; achraf.atila@gmail.com}

\affiliation{Federal Institute of Materials Research and Testing (BAM), Unter den Eichen 87, Berlin 12205, Germany}

\author{Yasser Bakhouch}
\affiliation{LS2ME, Facult\'{e} Polydisciplinaire Khouribga, Sultan Moulay Slimane University of Beni Mellal, B.P 145, 25000 Khouribga, Morocco}

\author{Zhuocheng Xie}
\affiliation{Institute of Physical Metallurgy and Materials Physics, RWTH Aachen University, 52056 Aachen, Germany}

\date{\today}

\begin{abstract}
Oxide glasses have proven to be useful across a wide range of technological applications. Nevertheless, their medium-range structure has remained elusive. Previous studies focused on the ring statistics as a metric for the medium-range structure, which, however, provides an incomplete picture of the glassy structure. Here, we use atomistic simulations and state-of-the-art topological analysis tools, namely persistent homology (PH), to analyze the medium-range structure of the archetypal oxide glass (Silica) at ambient temperatures and with varying pressures. PH presents an unbiased definition of loops and voids, providing an advantage over other methods for studying the structure and topology of complex materials, such as glasses, across multiple length scales. We captured subtle topological transitions in medium-range order and cavity distributions, providing new insights into glass structure. Our work provides a robust way for extracting the void distribution of oxide glasses based on persistent homology.
\end{abstract}

\maketitle

\section{Introduction}
The most recent definition of a glass states that a \textbf{\textit{glass is a nonequilibrium, noncrystalline condensed state of matter that exhibits a glass transition. The structure of glasses is similar to that of their parent supercooled liquids (SCL), and they spontaneously relax toward the SCL state. Their ultimate fate is to solidify, i.e., crystallize}}~\cite{Zanotto2017}. The noncrystalline nature of glasses gives them many advantages over crystalline materials, such as superior mechanical behavior and defiance of stoichiometry rules~\cite{Wondraczek2022, Rouxel2007, Rodrigues2022, Shakhgildyan2020, Kharouji2024}. However, this comes with drawbacks as well, as it is nontrivial to analyze the glass structure at a level larger than the short range ($>5\text{\AA}$)~\cite{Wondraczek2022, Kharouji2024, Srensen2022, Bakhouch2024, Atila2024b, Atila2024a, Atila2025}, thus resulting in a lack of proper structure–properties relationships.

Early theories of glass formation and structure were developed at the beginning of the twentieth century~\cite{tammann1903kristallisieren, randall193013, goldschmidt1925, atila2023thesis}.
In a seminal paper, “The Atomic Arrangement in Glass~\cite{Zachariasen1932},” Zachariasen showed that glasses and compositionally identical crystals share the same atomic interactions, leading him to propose that glass is a random network of atoms with a short-range order. The focus on oxide glasses, whose high melt viscosity traps extra internal energy versus the crystalline state, has led to the finding that ionic crystals prone to vitrification have open frameworks in which small, highly charged cations sit in oxygen-surrounded polyhedra that share corners with their neighbors~\cite{Zachariasen1932, atila2023thesis}.

With the absence of the long-range order~\cite{Wondraczek2022, Zachariasen1932, Zhang2024}, oxide glasses still exhibit a definable structure at short (below~5~\text{\AA}) and medium-range (5–20~\text{\AA}) levels~\cite{Brckner1970, Elliotta1991}. The short-range structure—quantified by coordination numbers derived from diffraction experiments~\cite{Swenson1995PRB, BUCHNER2014NOC, Crupi2015PRB, ABELDASILVEIRA2024NOC, Onodera2020NPGAsia, Hirata2024NPGAsia} or atomistic simulations~\cite{Atila2019a, Atila2019b, Ouldhnini2021, Atila2022}—extends into the connectivity of neighboring polyhedra and the formation of atomic clusters, which likewise can be characterized via both experimental~\cite{Crupi2015PRB, Onodera2020NPGAsia, Hirata2024NPGAsia} and simulations~\cite{Atila2019a, Atila2019b, Ouldhnini2021, Atila2022, Ge2024PRB, Urata2024PRM}.

The medium-range structure is characterized by rings of varying sizes that form and evolve with composition and processing~\cite{Atila2019a, Bakhouch2024, Atila2024a, Pan2024}, alongside voids or cavities arising from the change in the packing density, bond angles, and polyhedral connectivity. The size and distribution of these voids depend on network connectivity and the presence of modifiers (e.g., Na, Ca), which disrupt the network by creating non-bridging oxygens and thereby alter the number and size of void spaces~\cite{Crupi2015PRB}.

Depending on the intended application, the presence of voids in oxide glasses may be either essential or undesirable. For example, voids are necessary in porous glasses that serve as filters for adsorbing gases, heavy metals, and organic compounds~\cite{Kim2024, Enke2016}. The interconnected voids and pores provide pathways for fluid flow or allow reactants to access catalytic sites. The existence of large voids in aerogels is highly desired for thermal insulation applications, where lightweight and low thermal conductivity are needed, such as in aerospace applications~\cite{Maleki2014}. This is because heat cannot be transferred efficiently in regions with voids~\cite{Amanifard2007}. On the other hand, voids are not desirable in structural applications or when enhanced mechanical properties are required, as they act as stress concentrators that lead to the initiation and propagation of cracks in the glass, ultimately resulting in catastrophic failure~\cite{atila2023thesis, Wondraczek2022, Kurkjian2010, Muralidharan2007}. The tensile strength of defect-free silica glass fibers has been shown to nearly reach the theoretical strength~\cite{Kurkjian2010, Bartenev1973}. Thus, having a robust method to detect and visualize these voids or cavities is necessary for a complete understanding of the structure–properties relationship, which is of utmost importance for the further development of oxide glasses with adapted properties and mechanical performance.

Compared to other structural building blocks of glasses (e.g., polyhedra and rings), there are only a few studies investigating the voids in oxide glasses~\cite{Crupi2015PRB, Jena2023, Malavasi2006, Muralidharan2007, Pedone2008ChemMat, Yang2021geo, Singh2025CTC}. For instance, Crupi et al.~\cite{Crupi2015PRB} used neutron diffraction to investigate the effect of composition and pressure on the first sharp diffraction peak (FSDP) in alkali borate glasses. They linked the origin of the FSDP to the periodicity of void boundaries in the glass's random network. Malavasi et al.~\cite{Malavasi2006} used Delaunay triangulation and Voronoi diagrams to investigate voids in models of silica glass. They also computed the solubility of noble gases in the glass and compared the obtained values to known experimental data. Moreover, they leveraged their method to investigate the fracture of silica glass during tensile deformation~\cite{Pedone2008ChemMat}.

Although similar distributions are reported, the results also show slight deviations from one another~\cite {Yang2021geo}. Thus, there is a need for a robust method for detecting voids and cavities. Recently, persistence homology (PH) has emerged as an effective tool for studying the topology of materials at different length scales~\cite{Hiraoka2016, Murakami2019, Srensen2020, Wang2025, Fang2025}. PH has been applied to analyze the topological properties of a wide range of materials, including disordered systems, nanomaterials, and polymeric solutions. It is used to characterize geometric features such as holes, voids, and connectivity, which are essential for understanding material behaviors. This method has advantages over others, as it provides an almost unbiased detection of the glass topology. 
It has been used to clarify the real-space origin of the first sharp diffraction peak~\cite{Srensen2020}, the origin of the mixed alkali effect on ionic conductivity of silicate glasses~\cite{Onodera2019}. Persistent homology was also applied to analyze the topological signature of structural anisotropy in tensile-strained silica and sodium silicate glasses~\cite{Pan2024}. Under extreme pressures, combining persistent homology with molecular dynamics simulations revealed compression-induced structural changes—namely the emergence of tetraclusters and increased atomic packing fraction~\cite{Murakami2019}. At very high pressures, the resulting topology resembled that of a pyrite-type crystalline phase, despite tetracluster formation being prohibited in the crystal; this distinction underscores the glass’s tolerance for distortions in oxygen clusters~\cite{Murakami2019}.

Here, state-of-the-art analysis tools, namely persistence homology (PH), are leveraged to analyze the ring-size distribution and the void-size distribution, providing insight into the location and distribution of voids in the glass network. Hydrostatic compression is used solely to perturb the glass network topology, generating a series of distinct medium-range structures. This allows for a rigorous validation of the method, independent of any specific pressure-driven change of properties. The results of PH reveal hierarchical topological shifts, loop collapse, and cavity annihilation in compressed silica glass, which are influential factors in studying material behavior, such as mechanical and transport properties.

\section{\label{Sec:Method}Methods}
\subsection{Potential model}
The interactions between atoms are modeled using the SHIK potential developed by Sundarararaman~\etal~\cite{Sundararaman2020} and yield good structural, mechanical, and vibrational properties when compared to experimental data~\cite{Sundararaman2020, Atila2024b}. In this model, atoms are treated as fixed charged points, with the charge of oxygen being composition-dependent, and interact via a short-range Buckingham potential and long-range Coulomb interactions. An additional repulsive r$^{-24}$ term was added to properly treat the interactions at high temperature and pressure~\cite{Sundararaman2020}. The short-range cutoff was set to 8.0~\text{\AA}, and the long-range interactions were solved using the damped-shifted-force model~\cite{Fennell2006} with 0.2~\text{\AA}$^{-1}$ as the damping coefficient and 10.0~\text{\AA} as the long-range cutoff. Potential parameters and partial charges can be found in Ref.~\cite{Sundararaman2020}.

\subsection{Glass preparation and cold compression}

Silica glass (SiO$_2$) was prepared using the melt-quenching technique, a method commonly employed for preparing bulk glasses in atomistic simulations~\cite{Ganisetti2023}. Initially, 20001 atoms were randomly inserted in a cubic simulation box with periodic boundary conditions imposed in all directions. Unless otherwise stated, the time step was set to 1~fs in all simulations. The samples were equilibrated at a high temperature (T = 5000 K) in the canonical ensemble (NVT constant number of atoms, volume, and temperature) for 100~ps. Then, the system was further equilibrated in the NPT ensemble (constant number of atoms, pressure, and temperature) for 1000~ps at the same temperature, with an imposed external pressure of 100 MPa, which was ramped down to 0 MPa during cooling. The equilibrated liquids were quenched from 5000 K to 300 K using a cooling rate of 1 K/ps. After cooling, a subsequent run in the NPT ensemble for 1000~ps was performed at 300 K and 0~MPa. All simulations were performed using the LAMMPS code~\cite{Thompson2022}, and all atomic visualizations were done using OVITO~\cite{Stukowski2009}.

The glass was subjected to a cold compression at 300 K by increasing the hydrostatic pressure from 0 to 100 GPa for 10 ns. The temperature and pressure were controlled using Nosé-Hoover chain thermostat~\cite{Nos1984, Hoover1985} and a Parrinello-Rahman barostat~\cite{Parrinello1981}. The structure factor of the obtained glasses at different pressures was calculated and compared to available experimental data (See Fig.~\ref{FIGS:sqn} in the supplementary materials, SM). The positions of almost all peaks in the simulated structure factors are in good agreement with those obtained experimentally, which serves as validation of the structures obtained using this model of silica glass.

\begin{figure*}[ht!]
\centering
\includegraphics[width=0.8\textwidth]{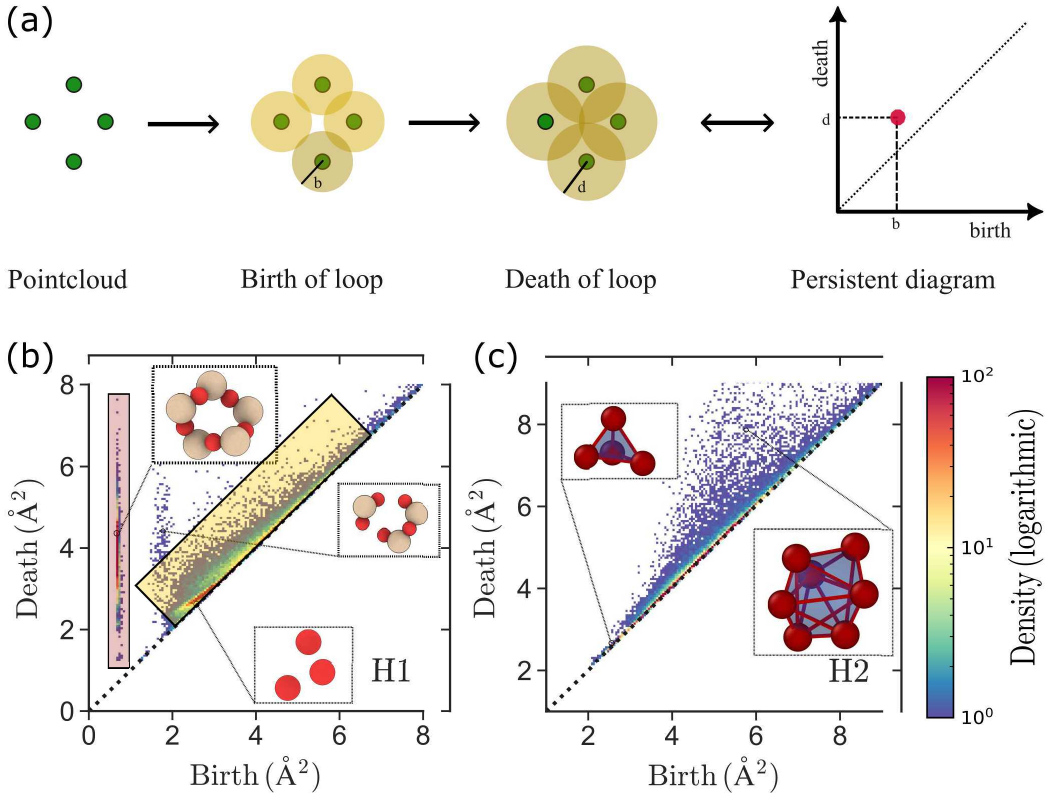}
\caption{(a) Schematic representation of the steps involved in the persistence diagrams construction. The steps involve having cloud points (left) and increasing of spheres' radii. When the spheres touch and make a closed loop, the coordinates of the radius at which this event happens are recorded as the birth time. With further increase of the radius, this loop will eventually close, and the radius at which this happens is recorded again as the death time of the loop. Both these birth and death pairs constitute a point on the persistence diagram (right), which is a histogram counting the number of voids on the birth-death plane. Persistence diagrams that correspond to (b) $H_1$ and (c) $H_2$ showing the topological feature of silica glass at 300 K and with 0 external pressure. The region highlighted by the transparent red box is for an island made of chemically connected loops, referred to later as rings as well. While the region highlighted in the yellow box is for the detected loops that are not chemically bonded.
The insets in (b) and (c) show snapshots of loops and cavities detected in the glass. The silicon and oxygen atoms are colored in light brown and red, respectively.}
\label{fig:pdc}
\end{figure*}

\subsection{Persistent Homology}
The topology of the glass was analyzed within the framework of persistent homology using Homcloud~\cite{Obayashi2022}.
The atomic positions in the glasses were used as input, with atomic radii assigned to each atom type. Then, the persistence diagram (PD) was constructed by progressively increasing the radii of each atom, with the assumption that each atom is a sphere, as schematically shown in Fig.~\ref{fig:pdc}(a). When the atoms touch each other in a closed loop, the radius at which this occurs is recorded and is referred to as the birth time. When the radii of the atoms increase and the loops disappear, that radius is recorded again and is referred to as the death time of the loop. This defines a pair of birth–death coordinates of the loop. The same analogy applies to cavities. Further details about persistent homology can be found elsewhere~\cite{Hiraoka2016, Obayashi2022, Srensen2022, Xiao2023, Wang2025}.

The process described above is used to construct a persistence diagram, a 2D histogram that represents the birth and death of topological features. These persistence diagrams $H_n$ are categorized by the dimension of the features extracted: $n = 0$ for clusters, $n = 1$ for loops, and $n = 2$ for cavities (see Fig.~\ref{fig:pdc}(b and c)).

\subsection{Analysis of the persistence diagrams}
Several topological features can be extracted from the persistence diagrams, such as loops and cavities. By analyzing the one-dimensional persistence diagram, we probe the topology of holes or loops that exist in the glass. Only the chemically bonded rings are of interest, which are found in a separated island at 0 GPa (see Fig.~\ref{fig:pdc}a). An initial radius of 0 was chosen for both Si and O atoms, which is nonphysical but allows for a clear separation between the chemically bonded loops and the other loops. This means that only the loops that have an equal number of Si and O atoms are considered to build the ring size distribution. The analysis was also performed using atomic radii extracted from the radial distribution functions, and the corresponding persistence diagrams are shown in the supplementary materials (SM, Figs.~\ref {figs:pd1_0} - \ref{figs:pd2_Si_r}).

A similar approach was used to extract the void-size distribution, where the two-dimensional persistence diagrams were analyzed to extract only physical voids. By this, we mean that voids inside a tetrahedron are not considered, and voids with substructure are also excluded (see inset of Fig.~\ref{fig:pdc}b).

\section{\label{Sec:Results}Results}
\subsection{Glass structure evolution with pressure}
\begin{figure}[h!]
\centering
\includegraphics[width=\columnwidth]{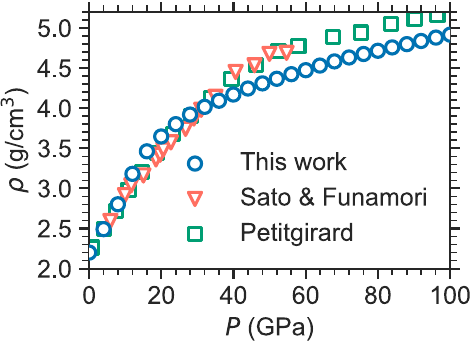}
\caption{The density change during the cold compression at 300 K compared with the experimental data from Refs.~\cite{Sato2008, Petitgirard2017}.}
\label{fig:dvp}
\end{figure}

The effect of the hydrostatic pressure on the glass was first assessed by tracking the evolution of the density with pressure and comparing it to previous experimental studies~\cite{Sato2008, Petitgirard2017}. Figure~\ref{fig:dvp} shows the density change with pressure up to 100 GPa and at room temperature. Initially, the density increases steeply with pressure. This increase is almost linear up to a pressure of around 18 GPa, followed by a curvature that spans between 18 GPa and a pressure of around 50 GPa. Finally, the density continues to increase, albeit at a lower rate than before. The rate of the change in density is almost identical to experimental measurements~\cite{Sato2008, Petitgirard2017}.

\begin{figure}[h!]
\centering
\includegraphics[width=\columnwidth]{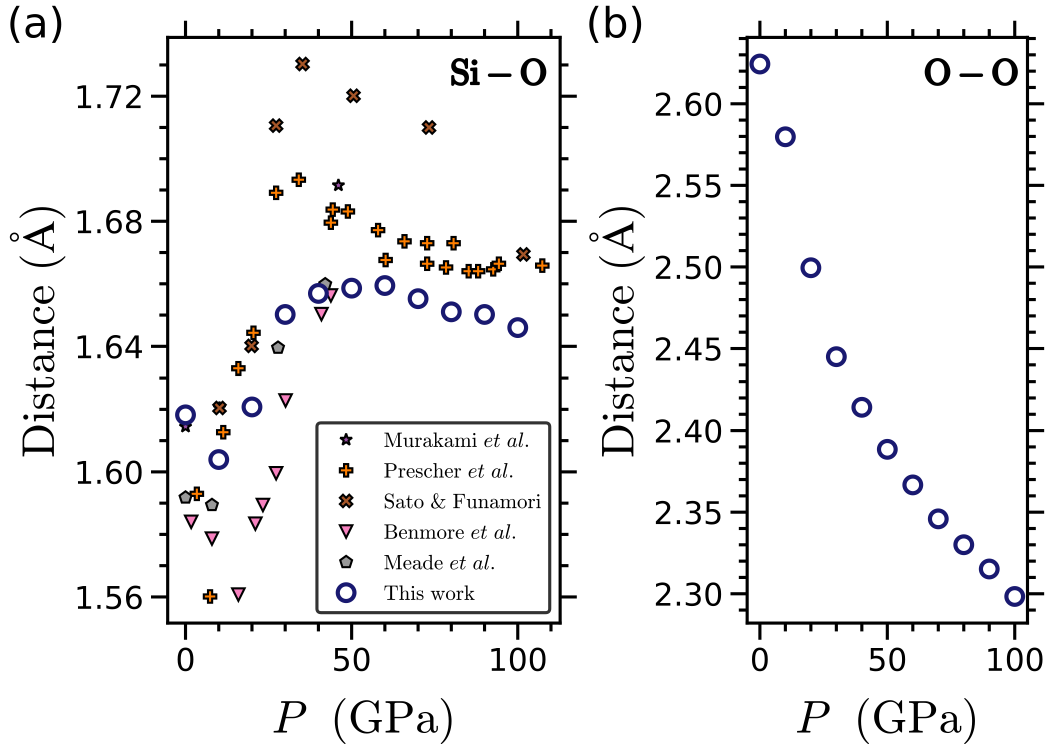}
\caption{Pair distances as a function of hydrostatic pressure (a) Si--O and (b) O--O calculated by fitting the first peak of the pair distribution functions (See Fig.~\ref{FIGS:rdf} in SM) to a skewed normal distribution~\cite{Sukhomlinov2017}. Experimental values for the Si--O pair distances are also given as a reference~\cite{Murakami2019, Prescher2017, Sato2010, Benmore2010, Meade1992}.}
\label{fig:BL}
\end{figure}

In Fig.~\ref{fig:BL}, we show data for the Si--O bond length, which was obtained by fitting the first peak of the radial distribution function to a skewed normal distribution~\cite{Sukhomlinov2017, Bakhouch2024}. Data from previous experimental works~\cite {Murakami2019, Prescher2017, Sato2010, Benmore2010, Meade1992} are also included in the same plot. A very good agreement is observed between the current simulation data and previous experiments.
The Si--O bond length initially decreases with increasing pressure up to around 10 GPa and then increases. At a pressure of around 60 GPa, the bond length begins to decrease again, highlighting three distinct domains of change as a function of pressure. This indicates the presence of different densification mechanisms, where different changes in the structure occur. The three domains can be explained by the changes that occur to the structure during densification and will be further discussed at different structural levels in the following sections. On the other hand, the O--O separation distance only decreased with increasing pressure, which is a sign of the change and distortion of the silicon-centered polyhedra.

\begin{figure}[h!]
\centering
\includegraphics[width=\columnwidth]{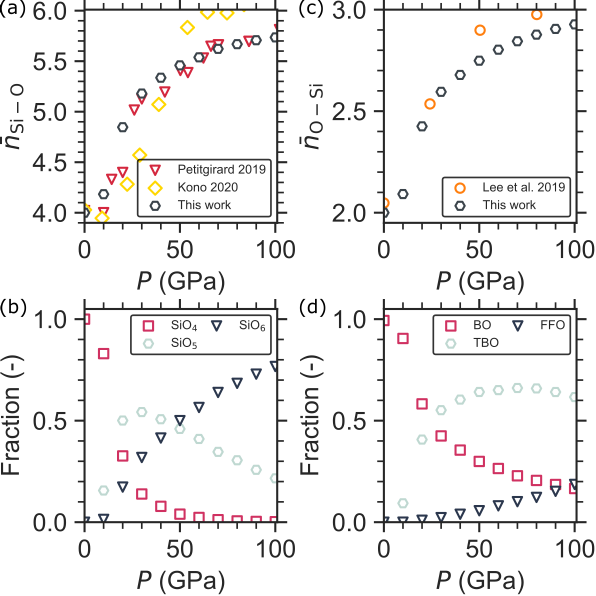}
\caption{Pressure dependence of (a) mean Si--O coordination number SiO$_n$, (b) the SiO$_n$ species, (c) mean O--Si coordination number, and (d) oxygen species in the glasses during compression. The average experimental data of the CN of Si--O in (a), and CN of O--Si in (c) are from Refs.~\cite{Kono2020, Petitgirard2019, Lee2019}.}
\label{fig:CN}
\end{figure}

By analyzing the coordination numbers of Si and O, one can gain further insights into the short-range structural changes that lead to the distinct densification regime of silica glass. Figure~\ref{fig:CN}(a) shows the mean coordination number of Si from our simulations and available experimental data~\cite{Kono2020, Petitgirard2019, Lee2019}, which is nothing more than the average number of O atoms in the first coordination shell of Si atoms. The mean Si-O coordination number increases with pressure and exhibits three distinct regimes, characterized by a change in the rate of increase. These changes in the Si--O coordination number are rationalized by probing the partial coordination number, which highlights that for each regime, a specific coordination number is dominant (see Fig.~\ref{fig:CN}(b)). For pressures below 18 GPa, fourfold coordinated Si atoms are the dominant species in the network, and the number of five-fold and six-fold Si atoms also exists with a lesser percentage. With a further increase in pressure from 18 GPa to around 60 GPa, and entering the next regime, the four-fold Si decreases notably, and the five-fold coordinated Si becomes the dominant species, with a non-negligible fraction of six-fold Si. At pressures higher than 60 GPa, few to no fourfold Si atoms exist, and the sixfold coordinated Si atoms are the dominant type of polyhedra. 

On the other hand, the change in O--Si coordination number is also of interest. The mean coordination number of O--Si is plotted in Fig.~\ref{fig:CN}(c) alongside experimental values from Lee et al.~\cite{Lee2019}. The experiments and simulations agree very well. The changes of O--Si mean coordination number follow the same changes and highlight three regimes, which can be explained through the partial O-Si coordination numbers (Fig.~\ref{fig:CN}(d)). At pressures lower than 30 GPa, the structure is dominated by O coordinated by two Si atoms. From pressures above 30 GPa, three-coordinated O is the dominant type of O atoms in the glass. At very high pressure ($P > 70$ GPa), the four-fold coordinated O starts to be non-negligible in the system with a percentage reaching values of around 6-7~\%, which is consistent with previous experimental data~\cite{Lee2019}.

Since the silica glass is a fully connected network of Si tetrahedra at room temperature, there are no non-bridging oxygens (NBO) in the glass, which makes the average coordination number (CN) of Si--O equal to the network connectivity (NC) as well. The NC or CN increases from 4 to 5.8 with pressure for silica, due to the emergence of high fractions of Si$_5$ and Si$_6$ species.

\begin{figure*}[htb!]
\centering
\includegraphics[width=\textwidth]{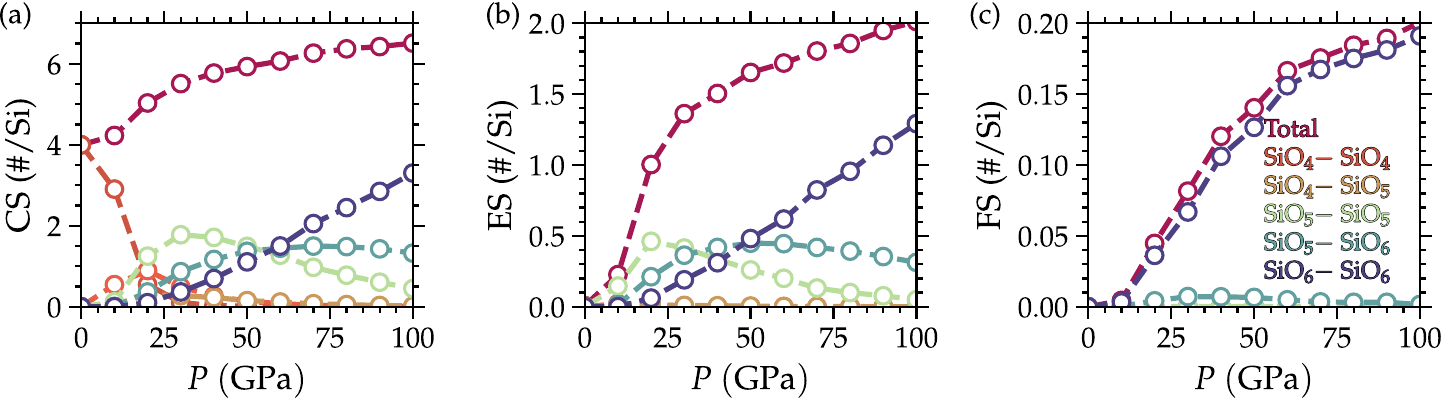}
\caption{SiO$_m$--SiO$_n$ sharing modes (a) Corner, (b) Edge, and (c) Face per polyhedron calculated as a function of pressure.}
\label{fig:connectivity}
\end{figure*}

To gain more insights into the change of the glass structure with densification, the connection modes between Si polyhedra are analyzed and categorized into three connection modes: corner-sharing, edge-sharing, and face-sharing. These sharing modes are categorized based on the number of BOs shared between adjacent polyhedra: corner-sharing (1 BO), edge-sharing (2 BOs), and face-sharing (more than 2 BOs).
Fig.~\ref{fig:connectivity} illustrates the connection modes between different SiO$_n$ polyhedra, which are corner-sharing (CS), edge-sharing (ES), and face-sharing (FS). The data is plotted to show the number of specific sharing modes per Si polyhedron. At ambient pressure, the network is made exclusively through corner-shared Si tetrahedra, which was expected from the values of the Si--O coordination numbers. As the pressure increases, the changes in the Si--O coordination numbers translate into changes in the connection modes between Si polyhedra from being simple corner-sharing to more complex behavior (See Fig.~\ref{fig:connectivity}). The number of corner-shared SiO$_4$--SiO$_4$ decreases and disappears at a pressure around 40 GPa. An observation worth mentioning is that the only connection mode between SiO$_4$--SiO$_4$ is corner sharing, even at elevated pressures. Other types of connection modes appear and disappear at different pressures, consistent with the abovementioned thresholds.
Moreover, another noticeable result is that the fraction of face-sharing between polyhedra is very low, accounting for less than 0.2 per Si polyhedron, and is almost all between SiO$_6$--SiO$_6$, which is consistent with previous \textit{ab initio} simulations~\cite{Hasmy2021}. 

\subsection{Glass topology}
To gain a deeper understanding of the changes that occur during densification, the topology of the glass was analyzed using persistence homology, as described in the Methods section. Key outputs of this method are the persistence diagrams, which are two-dimensional histograms that visualize various topological features such as clusters, loops/rings, and voids or cavities.

\begin{figure*}[htb!]
\centering
\includegraphics[width=0.8\textwidth]{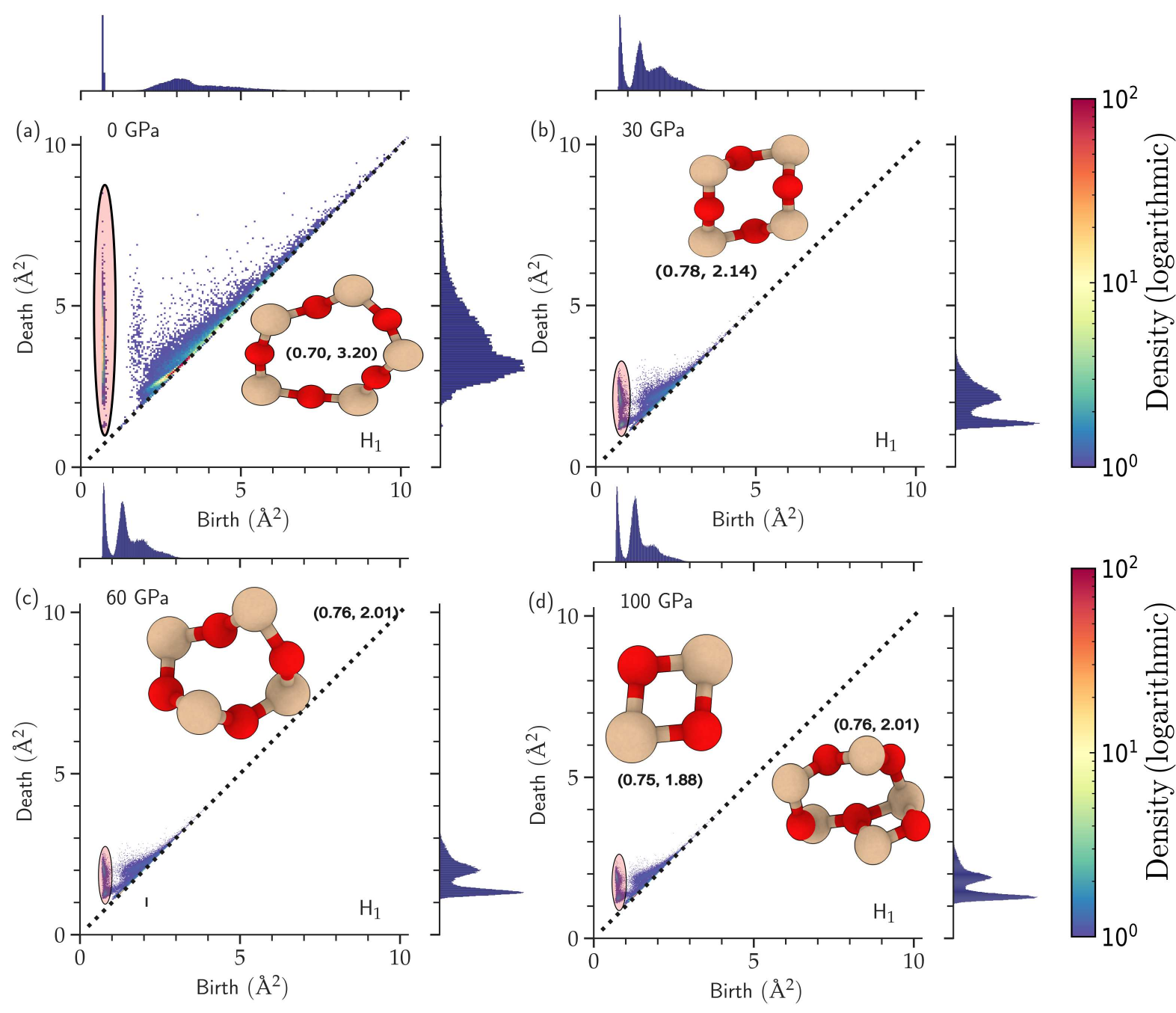}
\caption{One-dimensional persistence diagram of silica glass at (a) 0 GPa, (b) 30 GPa, (c) 60 GPa, and (d) 100 GPa. The insets show representative loops detected from the persistence diagrams with their birth-death coordinates, where Si atoms are shown in Brown and O atoms in Red. The persistence diagrams are calculated by setting the radius of both Si and O to $r = 0$~\text{\AA}. The elliptic region in red highlights the location of the chemically bonded loops.}
\label{fig:PDH1}
\end{figure*}

Figure~\ref{fig:PDH1} shows H$_1$ persistence diagrams of the silica glass, which describe one-dimensional topological features in the silica network, at different pressures during the densification. It is worth reminding that the dimensionality of the persistence diagrams is given in terms of Betti numbers, which are used to distinguish topological spaces based on the connectivity of $n$--dimensional simplicial complexes. In this notation, H$_0$, H$_1$, and H$_2$ are the persistence diagrams of connected components, one-dimensional loops, and two-dimensional "voids" or "cavities", respectively. 

\begin{figure*}[htb!]
\centering
\includegraphics[width=0.8\textwidth]{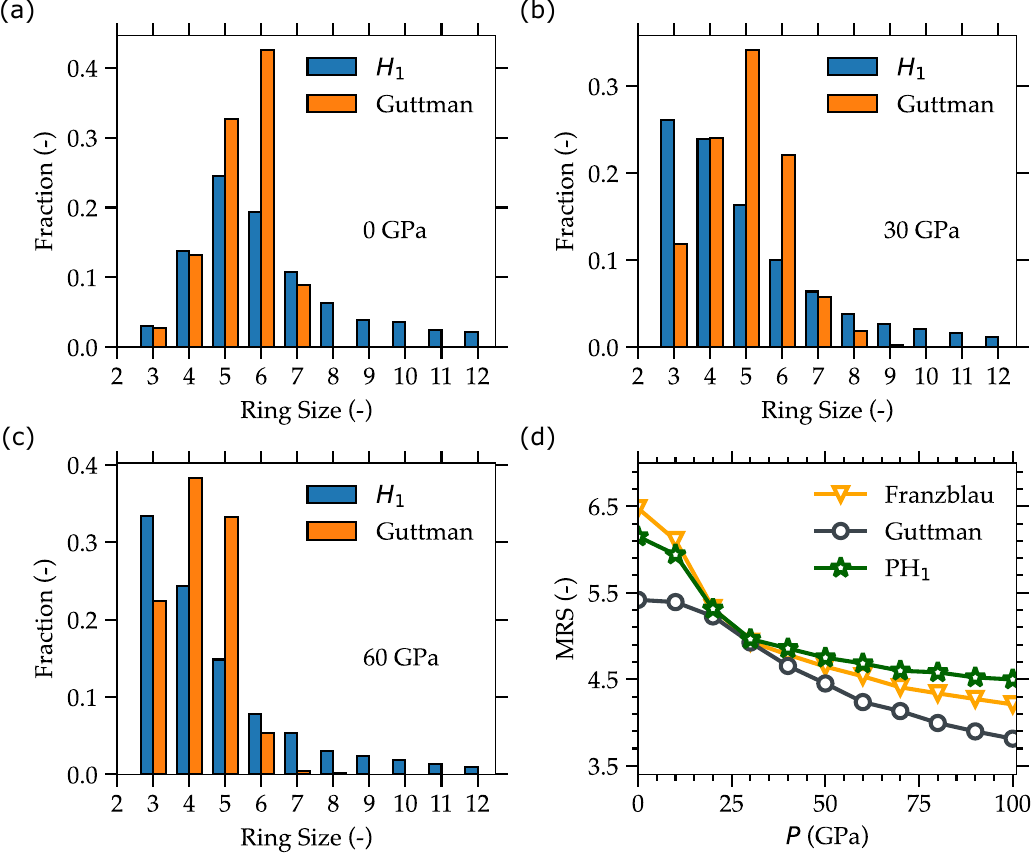},
\caption{Ring-size distribution obtained using Guttman criterion~\cite{Guttman1990} and the one obtained from persistence diagram $H_1$ from three selected pressures (See Fig.~\ref{FIGS:ringsall} in SM for the ring-size distributions at different pressures). (a) For the pressure-free silica glass model, (b) and (c) are for silica glass at pressures of 30 and 60 GPa, respectively.
(d) The mean ring size (MRS) variation as a function of pressure during compression.}
\label{fig:mrs}
\end{figure*}

At $P$ = 0 GPa, the persistence diagram shows two regions or groups of loops. The loops with the early birth time (around 1 \AA$^2$) are the ones with an equal number of Si and O, making --O--Si--O-- rings. This indicates that these loops are made through chemically bonded Si--O atoms. The other points on the H$_1$ PD, especially the ones along the diagonal, are for the loops that are formed by Si and O atoms that are not necessarily chemically bonded. 
The loops can be classified into two categories. The first category, which is found at short birth times, represents rings of interest, whose histograms are shown in Fig.~\ref{fig:mrs}. The others, which are located near the diagonal of the PD are mainly the loops found within those rings 'sub-loops'. 

With the increase of the pressure, both the histograms of birth times and death times of the loops shift towards the left, indicating the formation of fewer loops in general and small loops in particular. This significant change that occurs to the characteristic islands makes them coalesce and merge. We note that the "islands" refer to clusters of persistent features corresponding to the chemically bonded rings found at early birth times. In contrast, the "broad diagonal region" corresponds to transient loops formed by non-bonded atoms.
At the same time, new regions emerge, including an island near the death axis, accompanied by a region containing 2-membered loops indicative of edge-sharing between polyhedra and consistent with increased edge-sharing observed in Fig.~\ref{fig:connectivity}(b). 
Concurrently, the broad diagonal region differentiates into three distinct distributions: a region near the origin containing O--Si--O entities, a high-birth region primarily composed of three-membered oxygen loops, and an intermediate region with mixed loop compositions. The insets in Fig.~\ref{fig:PDH1} highlight the different types of rings found in the glass at different pressures, including the edge shared polyhedra, which are not considered as rings.

Figure~\ref{fig:mrs}(a - c) shows ring size distribution found from the H$_1$ PD and compared to the one calculated using the Guttman algorithm, as implemented in sovapy~\cite{Shiga2023}, at selected pressures. At $P = 0$ GPa, the ring-size distributions found from both methods are slightly different, where the rings extracted from the PD have larger sizes (8- and 9-membered rings) compared to the ones obtained from Guttman algorithm. The deviations persist at elevated pressures, and a noticeable increase of the population of small rings (3-membered rings) increases tremendously. This translates to a decrease of the mean ring size with pressure from values around 6 to values around 4 (See Fig.~\ref{fig:mrs}(d)). It is worth mentioning that the data of the mean ring size is consistent with the model suggested by Franzblau~\cite{Franzblau1991, Grigorev2024}.

\begin{figure*}[htb!]
\centering
\includegraphics[width=0.8\textwidth]{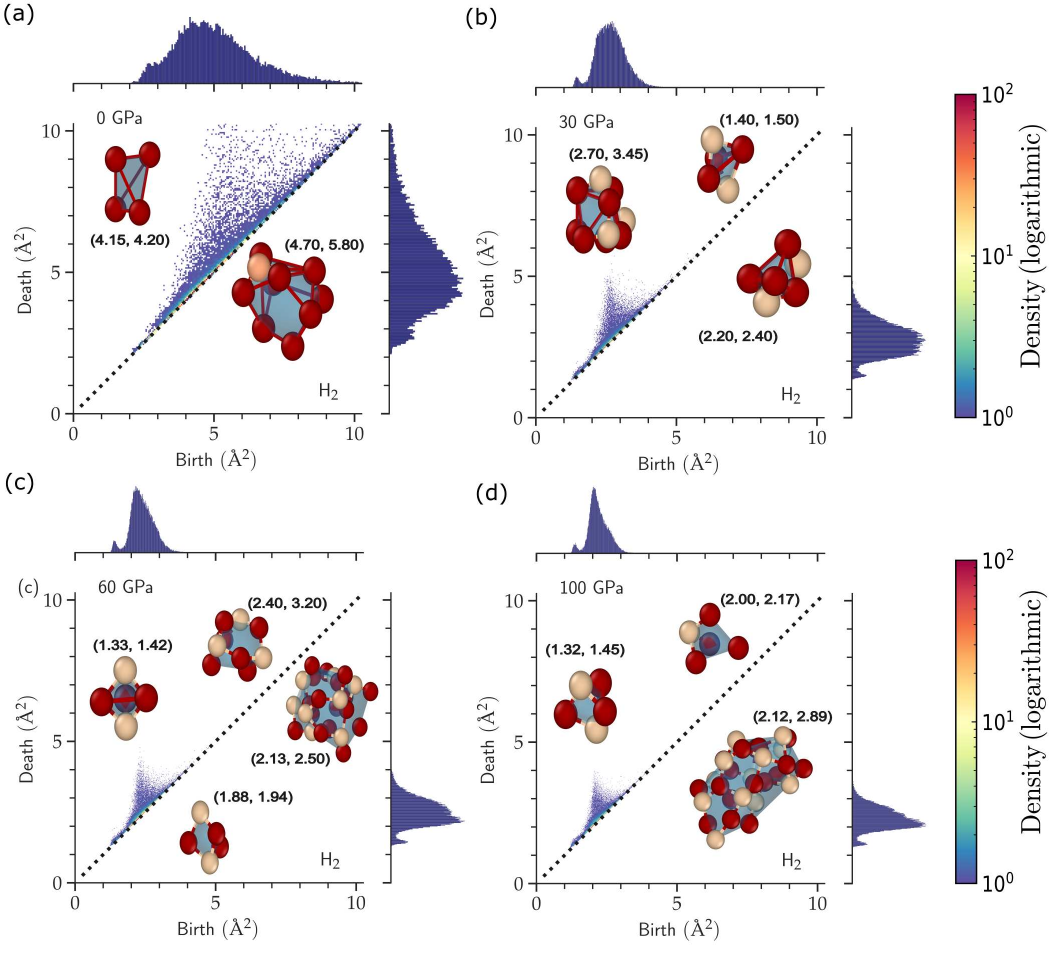},
\caption{Two-dimensional persistence diagram of silica glass at (a) 0 GPa, (b) 30 GPa, (c) 60 GPa, and (d) 100 GPa. The insets show representative enclosed voids detected from the persistence diagrams with their birth-death coordinates, Si atoms are shown in Brown and O atoms in Red. The persistence diagrams are calculated by setting the radius of both Si and O to $r = 0$~\text{\AA}. }
\label{fig:PDH2}
\end{figure*}

Since, at room temperature, the glass density is considerably lower than that of a SiO$_2$ crystal (e.g., $\alpha-$quartz has a density of 2.648 g/cm$^3$), there must be a type of free volume that exists in the glass. The latter is found in the form of voids or cavities that exist within the network. Similarly to the rings that we were able to extract from the one-dimensional persistence diagrams, the void size distribution will be extracted from the two-dimensional persistence diagrams at different pressures. 
The two-dimensional persistence diagrams H$_2$ for selected pressures are shown in Fig.~\ref{fig:PDH2}. These H$_2$ diagrams reveal details about the behavior of the interstitial space within the glass structure. 

At 0 GPa, the birth and death of the H$_2$ diagram show a broad distribution ranging from 2.5 to 10 $\text{\AA}^2$. A very high density is found along the diagonal, where voids or cavities that have a very small life times are found, with the lifetime $l$ of a cavity is defined as ($l = d_i - b_i)$, where $d_i$ is the death time of a cavity $i$ and $b_i$ is the birth time of the same cavity. These small lifetime-detected cavities are not physically meaningful voids. They are either voids within polyhedra or voids that are not formed through chemically bonded atoms and are therefore not considered in further analysis. As the pressure increases, the distribution of birth and death times becomes narrower and shifts towards smaller values, and the lifetime of the cavities also decreases. At 100 GPa, very small to almost no cavities exist anymore. This trend of the H$_2$ diagrams with pressure is consistent with the intuition that suggests that the voids will shrink and disappear with increasing pressure. Moreover, new distinct domains emerge with increasing pressure, corresponding to the different regions related to the face-sharing mode between polyhedra. 

\begin{figure}[!h]
\centering
\includegraphics[width=\columnwidth]{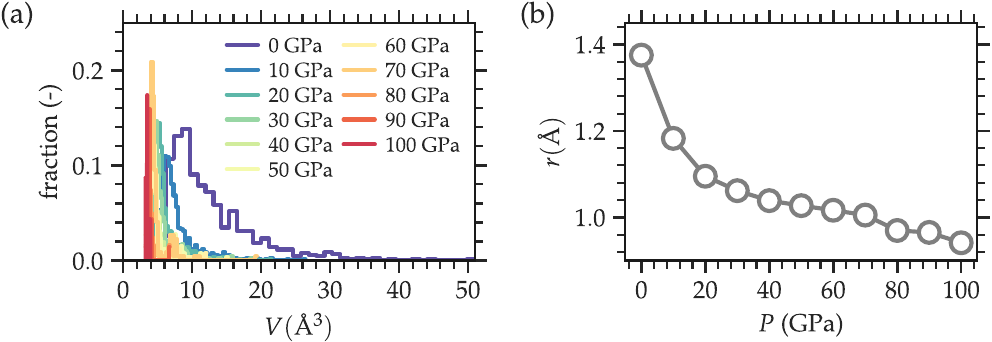}
\caption{(a) Voids-size distribution extracted from the H$_2$ diagrams. (b) The mean void radius as a function of the pressure.}
\label{fig:voidvolume}
\end{figure}

\begin{figure*}[t!]
\centering
\includegraphics[width=\linewidth]{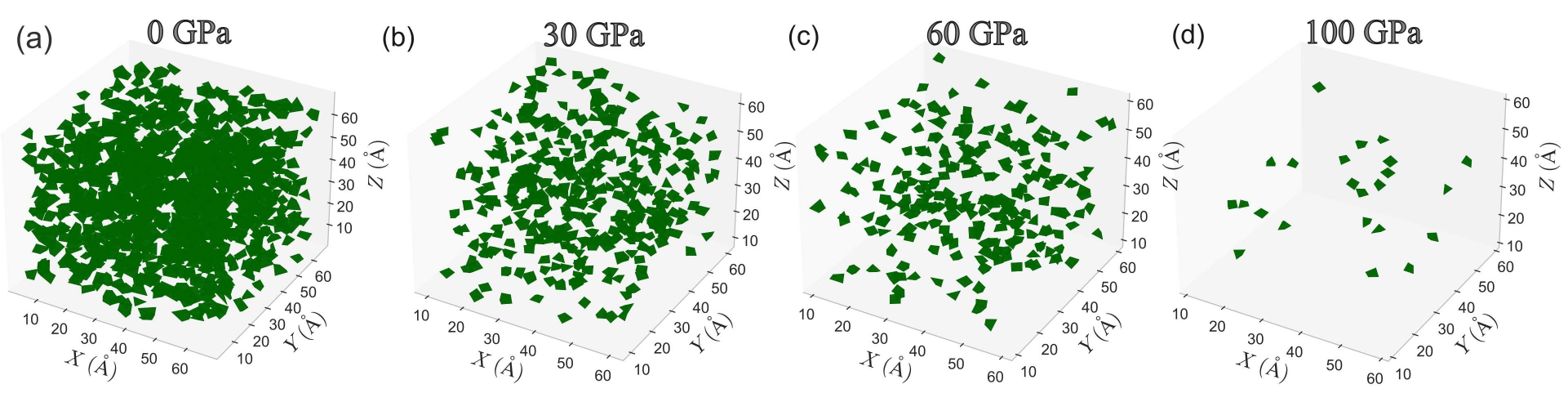}
\caption{Spatial distribution of the cavities detected from the H$_2$ diagrams at different pressures.
}
\label{fig:Void}
\end{figure*}

To further analyze the data from the H$_2$ diagrams, the coordinates of the atoms located at the edge of the detected cavities are extracted and filtered according to the criteria defined in the method section. It is worth emphasizing that we only analyzed the primitive voids, meaning that voids with child structures were not included in the analysis. This ensures the exclusion of both internal substructures, thereby isolating only interstitial spaces. Volumetric quantification of these voids was performed using the ConvexHull algorithm (scipy module)~\cite{2020SciPy-NMeth}, which determines the minimal convex polyhedron encompassing each void. A histogram of the void volume is then plotted in Fig.~\ref{fig:voidvolume}(a). The void volume for almost all pressures follows a skewed distribution with a long tail, indicating the existence of some relatively large voids in the structure. The obtained void size distribution is in fact in agreement with previous studies~\cite{Malavasi2006, Ono2012}. The differences that might be noticeable compared to previous papers come from the methods used, where the void size distribution found depends on a pre-defined criterion. However, the voids found using PH are topologically grounded voids that exist in the glass structure, and the filtering criterion is only used to remove nonphysical voids. The volume distribution shifts towards smaller volumes as the pressure increases and is almost represented by a delta function at very high pressures, indicating that very few voids are found at such pressures and can be ignored. Moreover, for better comparison with data found in the literature, where the void-size distribution is usually reported as a function of the void radius, we plotted in Fig.~\ref{fig:voidvolume}(b) the mean void radius at different pressures. As intuitively expected, it decreases with increasing pressure. The mean void radius was calculated from the volume distribution, assuming that the voids are spherical, although in reality, they are not. The spatial distribution of such voids is shown in Fig.~\ref{fig:Void} for four selected pressures and shows that the number of voids decreases with increasing pressure.

\section{\label{Sec:Discussion}Discussion}

The structure and topology of silica glass subjected to cold compression were investigated at short- and medium-range orders. The results were compared to experimental data when available and showed a very good agreement. Significant changes to the glass structure and topology were caused gradually by the pressure. A transition from tetrahedral to octahedral glass network was observed with increasing pressure, in which the network becomes dominated by six-fold coordinated Si atoms instead of four-fold coordinated Si atoms at low pressures, leading to an increase in network connectivity and compactness of the sample. This transition is accompanied by a change of the Si--O bond length, which decreases, then increases, and finally decreases again, which is in agreement with previous simulations~\cite{Ouldhnini2021, Bakhouch2024} and experiments~\cite{Murakami2019, Prescher2017, Sato2010, Benmore2010, Meade1992, Weigel2019, Kono2020}. The initial decrease of the bond length is due to the decreased size of SiO$_4$ tetrahedra, which is supported by the fact that the network is still made out of $\approx$ 85\% of SiO$_4$. The transition regime is characterized by an increase in the population of the SiO$_5$ and an increase in the Si--O bond length. The increased bond length is mainly due to the presence of SiO$_5$ and SiO$_6$ in the network, which have longer bond lengths than SiO$_4$. At even larger pressure where the structure is dominated by SiO$_6$ and SiO$_5$, the bond length decreases with pressure, which is due to the shrinkage of the SiO$_6$ and SiO$_5$ polyhedra~\cite{Zeidler2014, Stebbins2019, DellaValle1996}. 

The pressure-induced changes at the scale in which two tetrahedra are involved (near medium-range order). For pure silica at ambient conditions, only corners shared tetrahedra are expected, however, as the pressure increases there will be changes to the electronic structure of the atoms (although not directly accessible from our simulations) and the nature of bonding between Si and O will be affected, where it will become more ionic as was highlighted before by Hasmy et al.~\cite{Hasmy2021} using ab initio-based calculations of compressed silica glass. These changes in bonding enhance the higher coordination of Si atoms, which in turn lead to a change in the connectivity type between SiO$_n$ polyhedra, from corner to more complex edge sharing and face sharing. These changes at the short-range and polyhedral connectivity have consequences on the ring-size distribution. The ring-size distribution as plotted in Fig.~\ref{fig:mrs} was, as intuitively expected, shifting towards lower values as the large rings will collapse into smaller rings with the increase of the pressure. 

The size distribution of voids and their spatial distribution in the glass network are crucial aspects of the structure that are not heavily discussed in the literature, except for the few papers introduced in the introduction. The void-size distribution was calculated from the persistence diagrams by filtering all voids and considering only the physical voids, as described above. As intended, the void size distribution indeed decreases with an increase in pressure, until no statistically meaningful voids exist in the sample. This is a direct consequence of the densification and the compactness of the network. Moreover, the void radius calculated from the H$_2$ PDs is consistent with the void radius found experimentally using positron annihilation~\cite{Ono2012, Ono2018}, estimates from gas solubility experiments~\cite{Nakayama1990}, and previous MD simulations~\cite{Malavasi2006, Deng2021}. 

It is worth stressing that PH calculations for MRO analysis are sensitive to the initial weight parameters~\cite{Xiao2023, Firooz2024}. In our analysis of chemically involved loops, we made the nonphysical assumption of equal radii (0 \text{\AA}) for Si and O atoms, which resulted in an isolated island of chemically bonded loops (both reducible and irreducible) at 0 GPa, which made it easy to detect such island visually. However, even when choosing proper radii, as those obtained from the radial distribution function, the ring size distribution is not affected and can also be extracted. These PH-based descriptors (loops and cavity volume) are ideal features for machine learning algorithms, which can be utilized for high-throughput screening of multicomponent glasses with targeted medium-range structures for specific applications, such as energy storage.

\section{\label{Sec:Conclusion}Conclusions}

In conclusion, persistent homology was used to extract the void-size distribution using a model network of glass. The pressure was used to tailor the topology of the glass and to alter the voids in the network, providing a test for validating the method used.

The ring-size distribution was also extracted from PD diagrams, and a major change to the distribution of ring sizes is observed with varying pressures. With increasing pressure, the glass network becomes mainly made through small-ring sizes, which was observed by the reduction of the lifetime of the rings, suggesting that the densification of silica glass is dominated by ring size reduction, which could be general for other silicate glasses as well~\cite{Murakami2019}. 

Our analysis extended to H$_2$ diagrams to statistically determine the distribution of cavities in silica glass. This approach yields better results compared to conventional methods in the literature, which often lack standardization due to their reliance on inserted initial spheres for estimating void radius distributions and their inability to account for varying coordination numbers in ionic radii calculations. The persistent homology calculation circumvented these limitations, though at the cost of increased computational time, which can be improved with proper parallelization. Only cavities meeting specific criteria were selected for analysis and discussion. These cavities were the ones that we speculated about in a realistic glass model.  
This work not only validates the application of topological data analysis in materials science but also opens new avenues for investigating the relationship between topology and physical properties in glass systems under extreme conditions. This is enabled by our successful extraction of cavities from the persistence diagrams, complementing coordination numbers and ring statistics, which can be used to extract the void location, shape, and void-size distribution in other amorphous samples such as glassy metal-organic framework (g-MOF)~\cite{Ding2024} and allow for accelerated rational design of new amorphous materials. 

\section*{Declaration of competing interest}
The authors declare that they have no known competing financial interests or personal relationships that could have appeared to influence the work reported in this paper.

\clearpage

\onecolumngrid
\begin{center}
\large Supplementary Materials to: \\ Revealing the Void-Size Distribution of Silica Glass using Persistent Homology
\end{center}
\begin{center}
Achraf Atila$^{1*}$, Yasser Bakhouch$^2$, and Zhuocheng Xie$^{3}$ \\ \vspace{0.1cm}
$^1$Federal Institute of Materials Research and Testing (BAM), Unter den Eichen 87, Berlin 12205, Germany \\ \vspace{0.1cm}
$^2$LS2ME, Facult\'{e} Polydisciplinaire Khouribga, Sultan Moulay Slimane University of Beni Mellal, B.P 145, 25000 Khouribga, Morocco \\ \vspace{0.1cm}
$^3$Institute of Physical Metallurgy and Materials Physics, RWTH Aachen University, 52056 Aachen, Germany \\ \vspace{0.1cm}

$^*$achraf.atila@bam.de; achraf.atila@gmail.com
\end{center}

\renewcommand{\thefigure}{S\arabic{figure}}
\renewcommand{\thetable}{S\arabic{table}}
\setcounter{figure}{0}
\setcounter{table}{0}

\maxdeadcycles=1000

\clearpage

\begin{figure}[ht]
\centering
\includegraphics[width=0.7\textwidth]{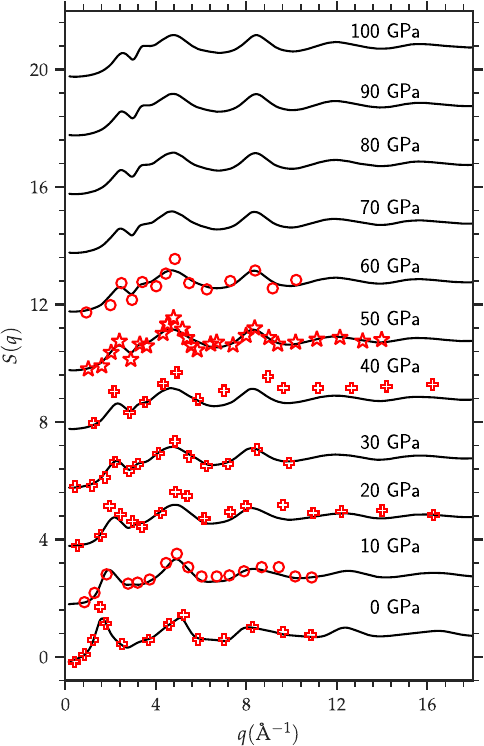}
\caption{The neutron structure factors of the simulated glasses at different pressures  (solid lines) compared to the experimental structure factors from Murakami et al.~\cite{Murakami2019} (circle symbols), Benmore et al.~\cite{Benmore2010} (plus symbols), and Sato et al.~\cite{Sato2010} (star symbols). The plots are shifted by one along the y-axis for clarity. }
\label{FIGS:sqn}
\end{figure}


\renewcommand{\arraystretch}{1}
\begin{table}
\centering
\caption{Atomic radii used in the construction of the persistence diagrams for glasses at different pressures. The radii are calculated based on the pair distribution function of Si-O and O-O at different pressures.}
\vspace{0.1cm}
\begin{tabular}{ccc}
\hline
\vspace{0.2cm}
Pressure (GPa) & {$r_\text{O}$} (\text{\AA}) & {$r_\text{Si}$} (\text{\AA}) \\
\hline
0  & 1.312 & 0.306 \\
10 & 1.289 & 0.314 \\
20 & 1.249 & 0.371 \\
30 & 1.222 & 0.427 \\
40 & 1.207 & 0.449 \\
50 & 1.194 & 0.464 \\
60 & 1.183 & 0.476 \\
70 & 1.172 & 0.482 \\
80 & 1.165 & 0.486 \\
90 & 1.157 & 0.492 \\
100 & 1.149 & 0.496 \\
\hline
\end{tabular}
\label{stab:tab1}
\end{table}


\begin{figure}
\centering
\includegraphics[width=\columnwidth]{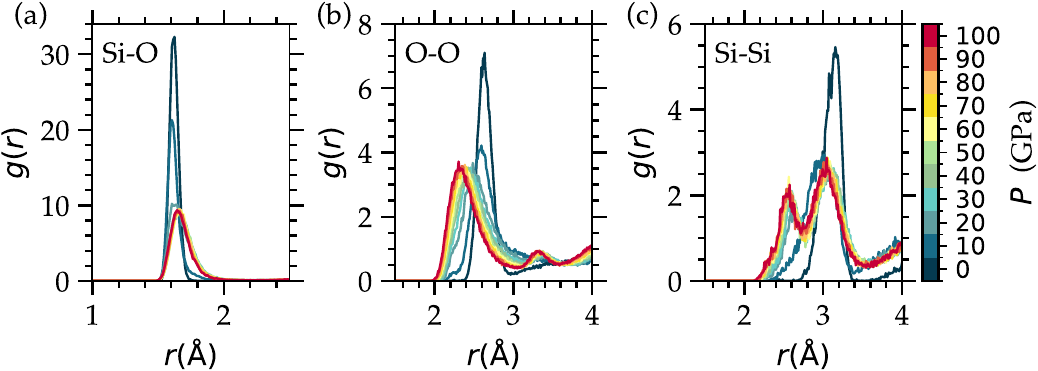}
\caption{Pair distribution functions of (a) Si--O, (b) O--O, and (c) Si--Si as a function of pressure.}
\label{FIGS:rdf}
\end{figure}

\begin{figure*}
\centering
\includegraphics[width=\textwidth]{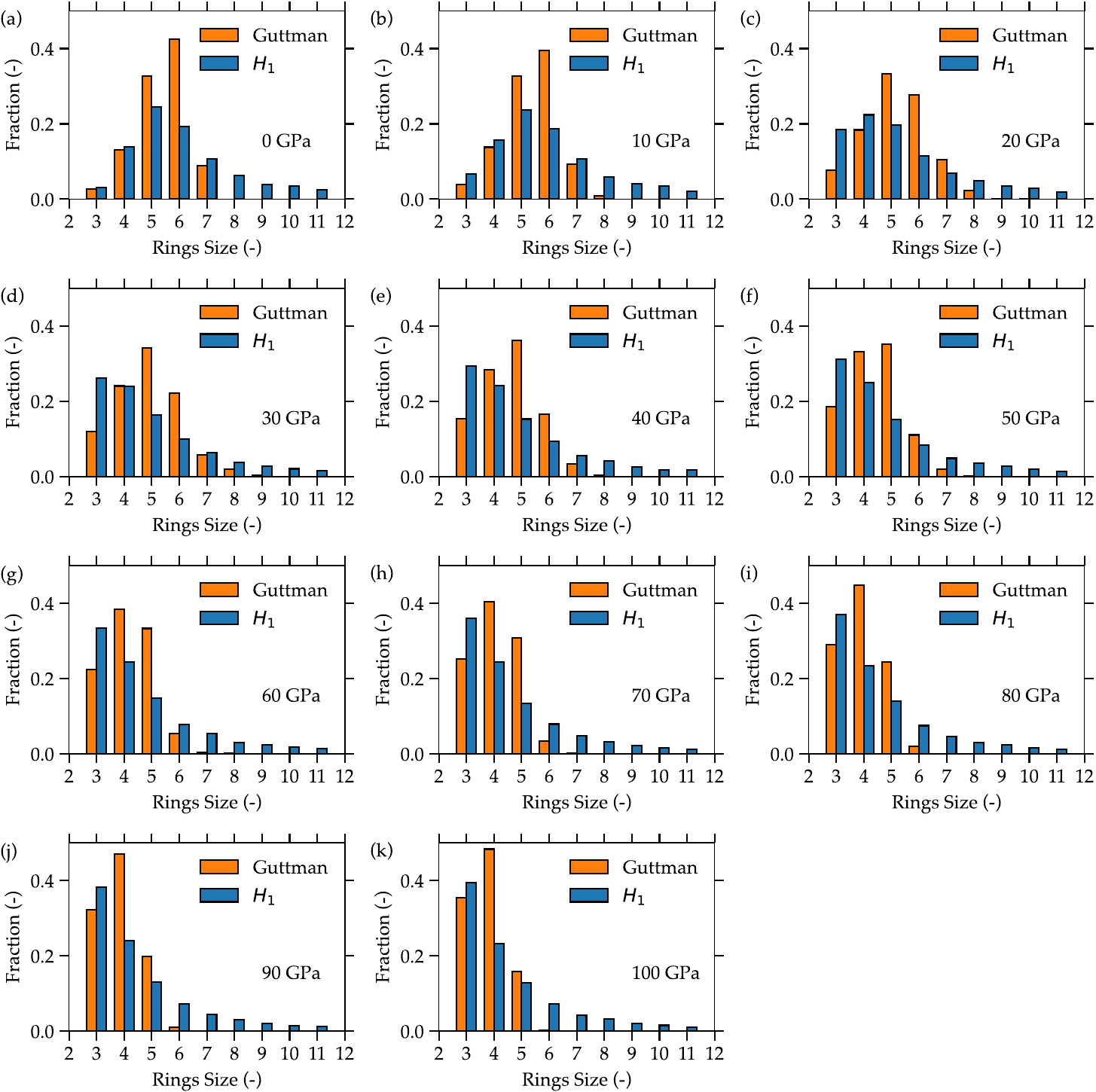}
\caption{Ring-size distribution obtained by Guttman criterion~\cite{Guttman1990} and the ones extracted from $H_1$ diagrams at different pressures during cold compression.}
\label{FIGS:ringsall}
\end{figure*}


\begin{figure*}[htb!]
\centering
\includegraphics[width=\textwidth]{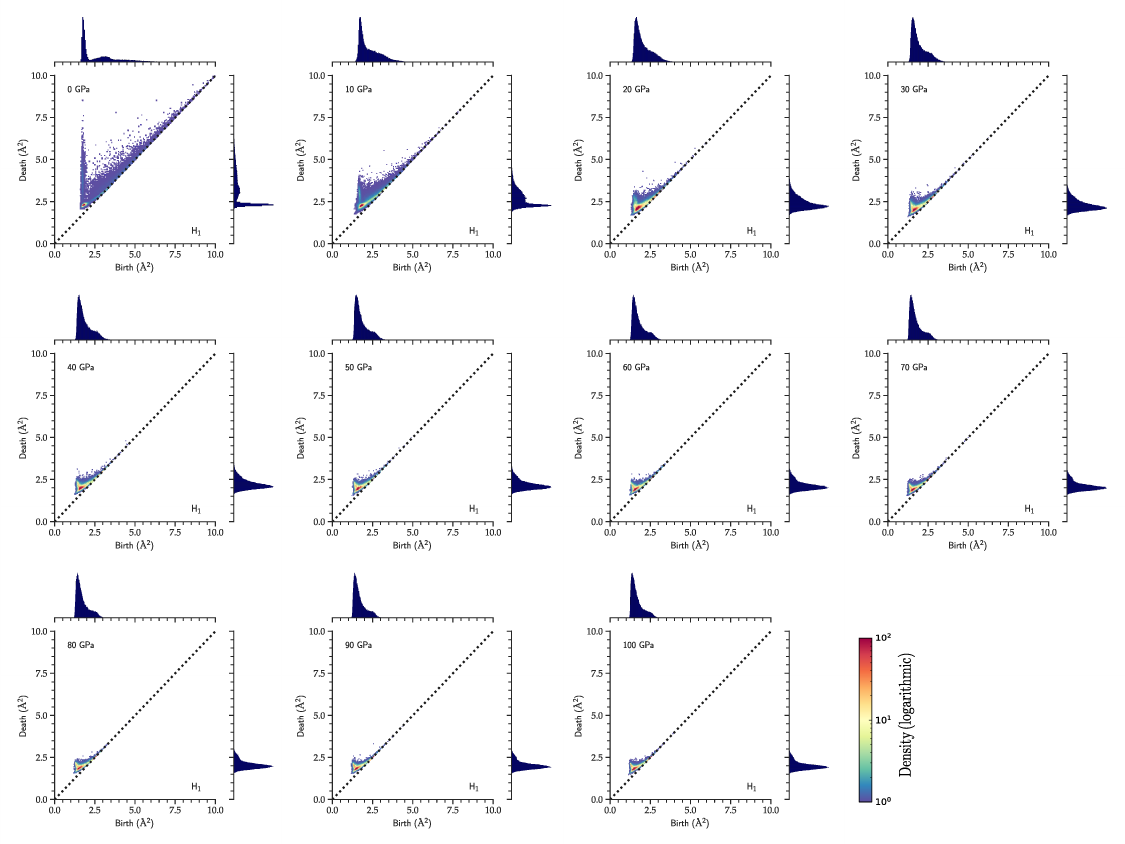}
\caption{One-dimensional persistence diagram of silica glass at different selected pressures along the compression trajectory from $P = 0$~GPa to $P = 100$~GPa at an increment of 10~GPa. The radius used for the Si and O atoms is $r = 0$~\text{\AA}.}
\label{figs:pd1_0}
\end{figure*}

\begin{figure*}[htb!]
\centering
\includegraphics[width=\textwidth]{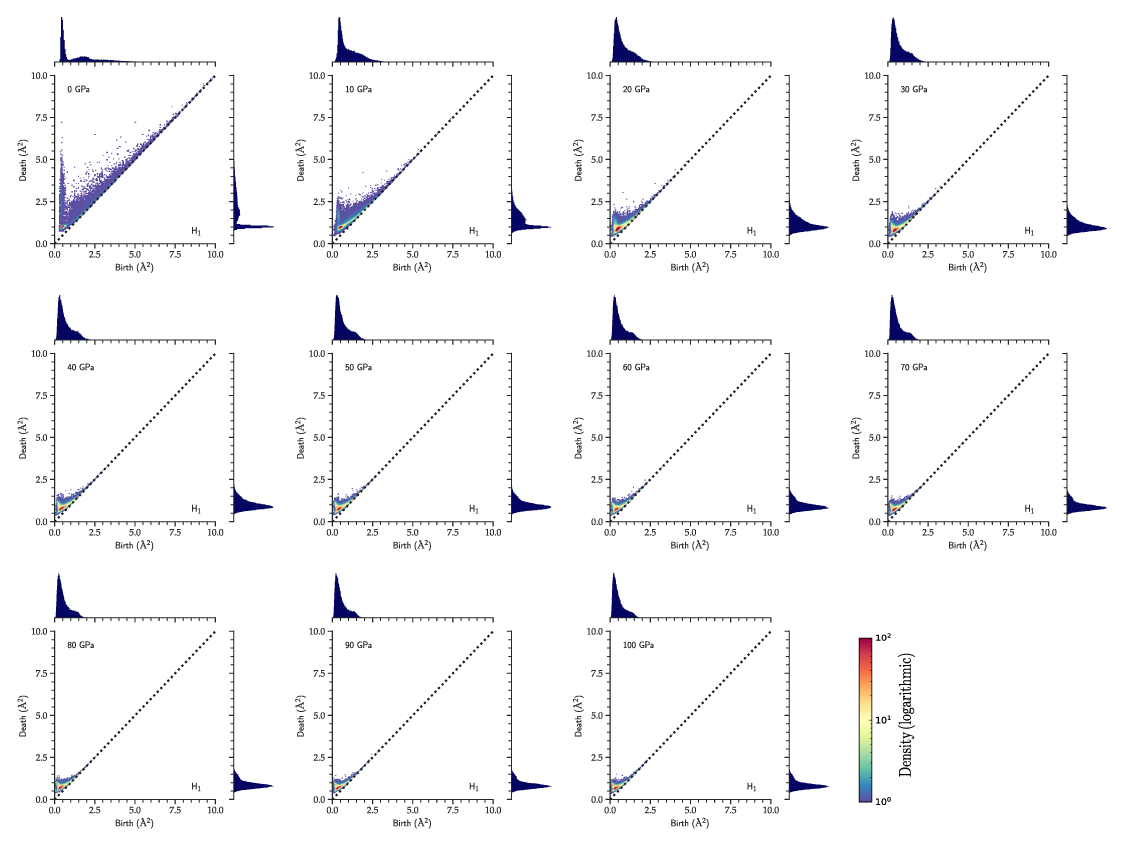}
\caption{One-dimensional persistence diagram of silica glass at different selected pressures along the compression trajectory from $P = 0$~GPa to $P = 100$~GPa at an increment of 10~GPa. The radii used for Si and O atoms are given in Tab.~\ref{stab:tab1}.}
\label{figs:pd1_r}
\end{figure*}

\begin{figure*}[htb!]
\centering
\includegraphics[width=\textwidth]{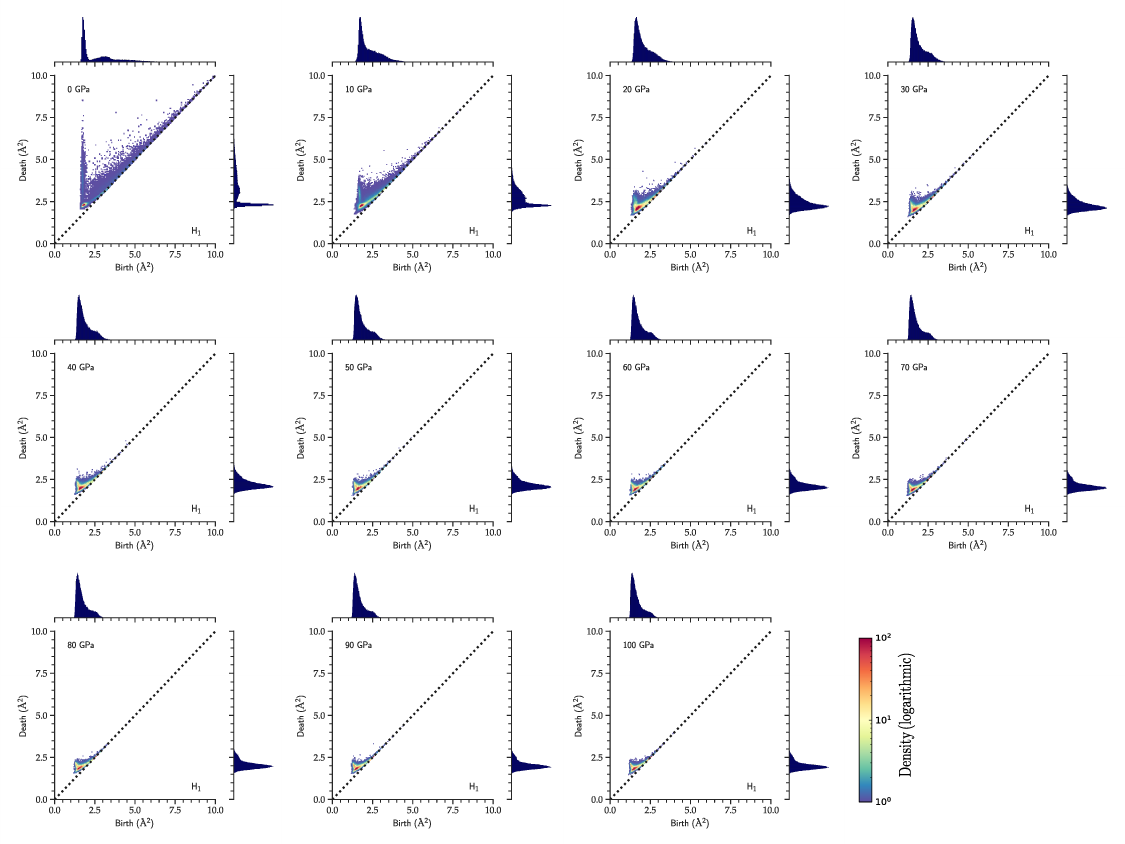}
\caption{O-centric one-dimensional persistence diagram of silica glass at different selected pressures along the compression trajectory for $P = 0$~GPa to $P = 100$~GPa at an increment of 10~GPa. The radius used for the O atoms is $r = 0$~\text{\AA}.}
\label{figs:pd1_O_r0}
\end{figure*}

\begin{figure*}[htb!]
\centering
\includegraphics[width=\textwidth]{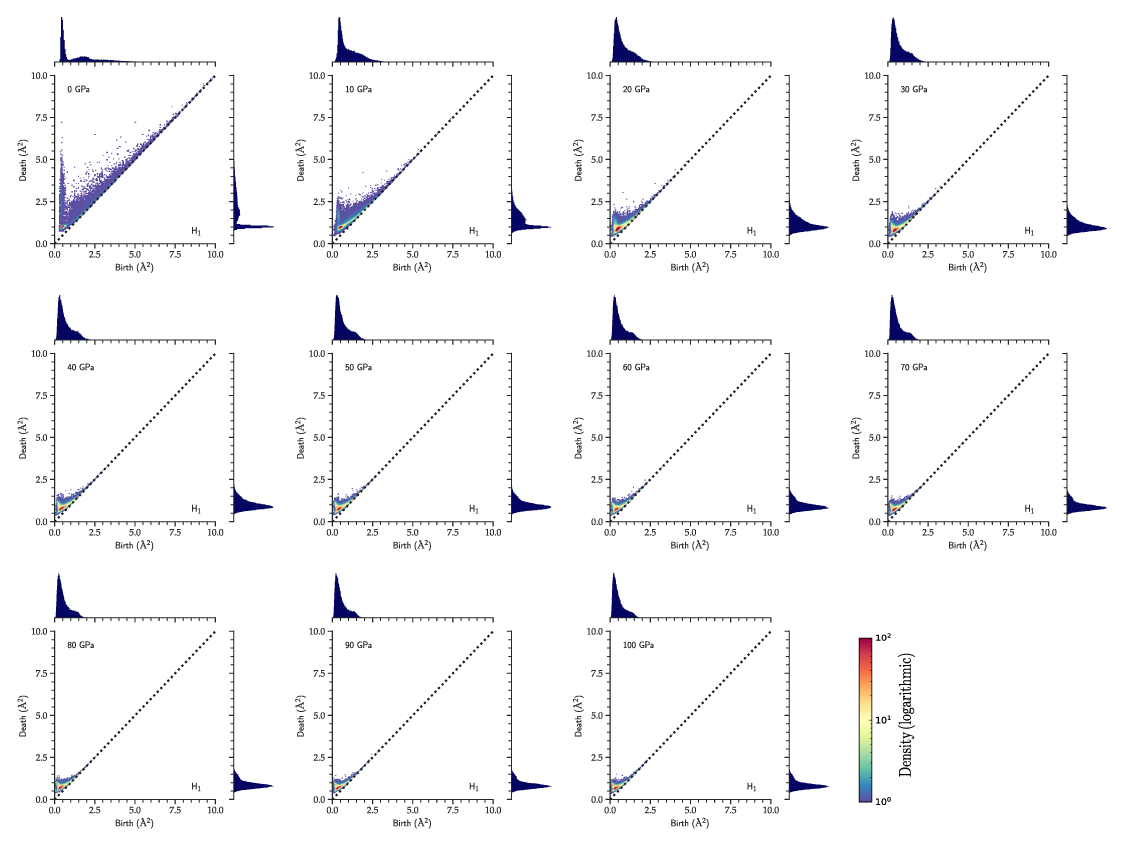}
\caption{O-centric one-dimensional persistence diagram of silica glass at different selected pressures along the compression trajectory for $P = 0$~GPa to $P = 100$~GPa at an increment of 10~GPa. The radii used for the O atoms are given in Tab.~\ref{stab:tab1}.}
\label{figs:pd1_O_r}
\end{figure*}

\begin{figure*}[htb!]
\centering
\includegraphics[width=\textwidth]{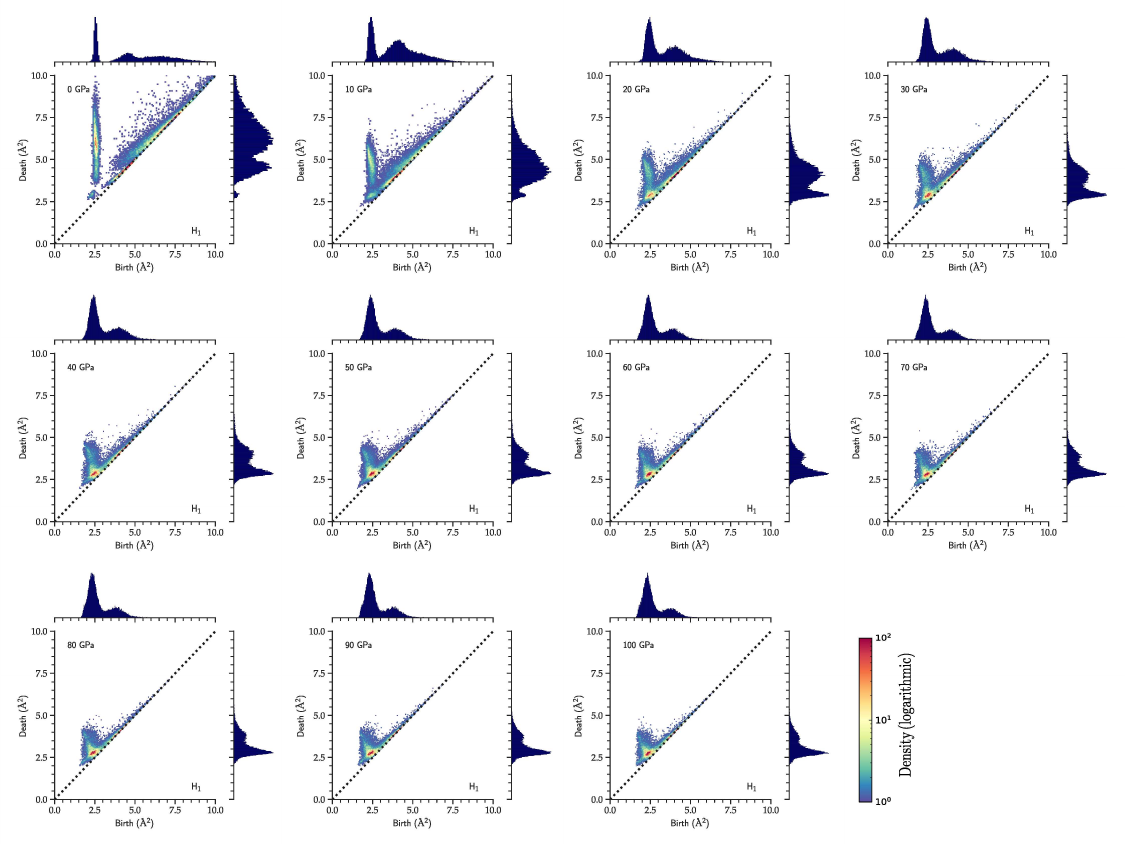}
\caption{Si-centric one-dimensional persistence diagram of silica glass at different selected pressures along the compression trajectory for $P = 0$~GPa to $P = 100$~GPa at an increment of 10~GPa. The radii used for the Si atoms are $r = 0$~\text{\AA}.}
\label{figs:pd1_Si_r0}
\end{figure*}

\begin{figure*}[htb!]
\centering
\includegraphics[width=\textwidth]{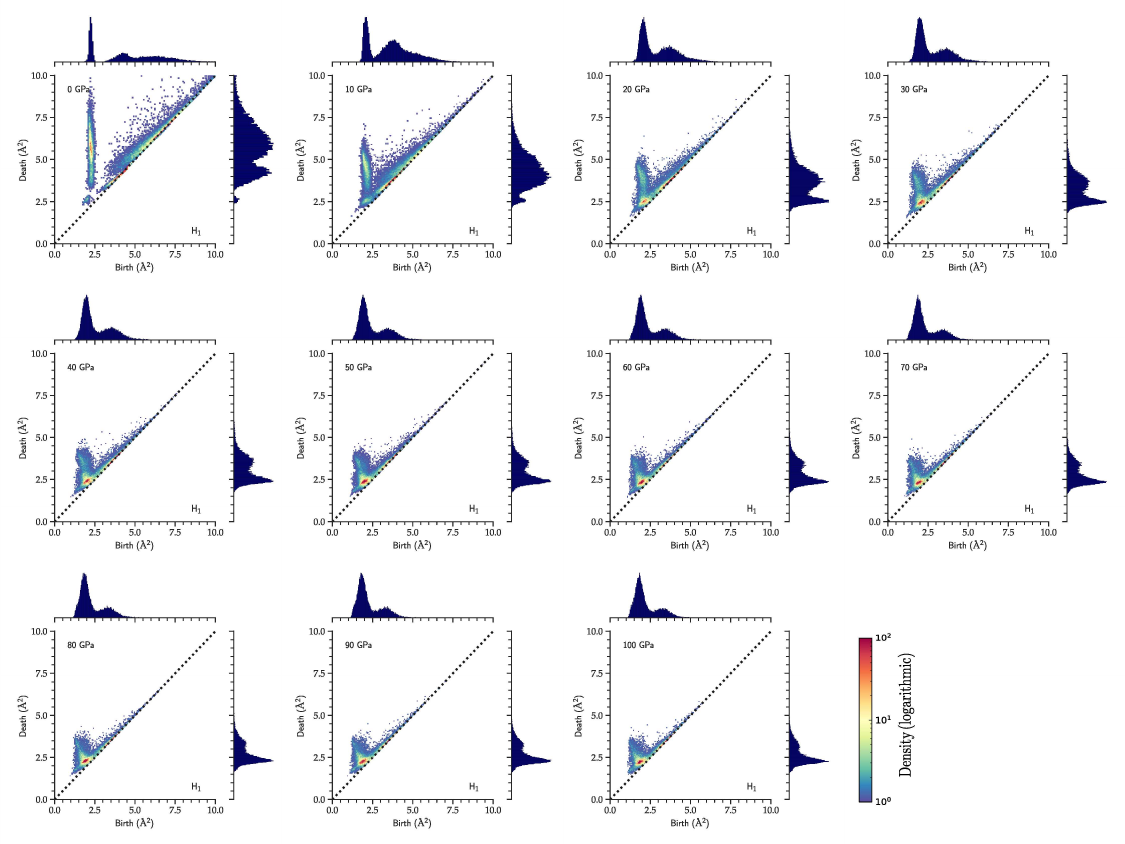}
\caption{Si-centric one-dimensional persistence diagram of silica glass at different selected pressures along the compression trajectory for $P = 0$~GPa to $P = 100$~GPa at an increment of 10~GPa. The radii used for the Si atoms are given in Tab.~\ref{stab:tab1}.}
\label{figs:pd1_Si_r}
\end{figure*}


\begin{figure*}[htb!]
\centering
\includegraphics[width=\textwidth]{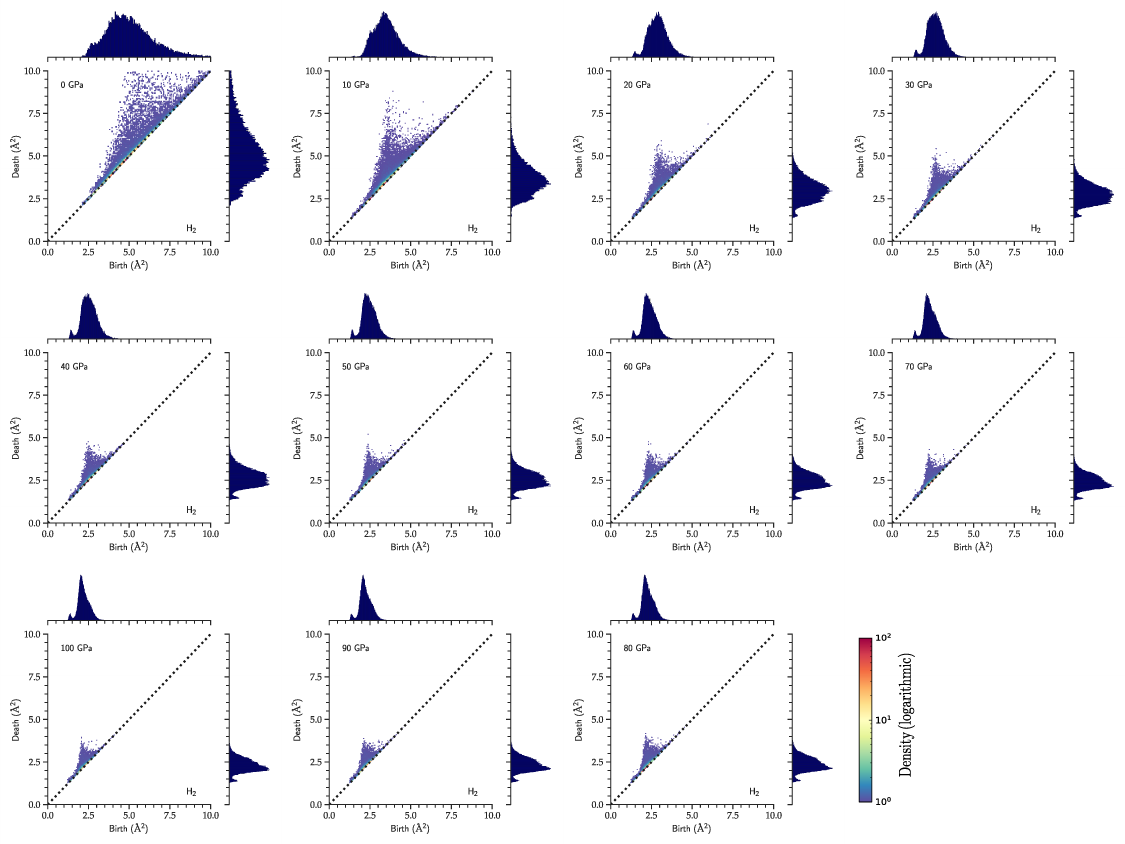}
\caption{Two-dimensional persistence diagram of silica glass at different selected pressures along the compression trajectory for $P = 0$~GPa to $P = 100$~GPa at an increment of 10~GPa. The radii used for the Si/O atoms are $r = 0$~\text{\AA}.}
\label{figs:pd2_r0}
\end{figure*}

\begin{figure*}[htb!]
\centering
\includegraphics[width=\textwidth]{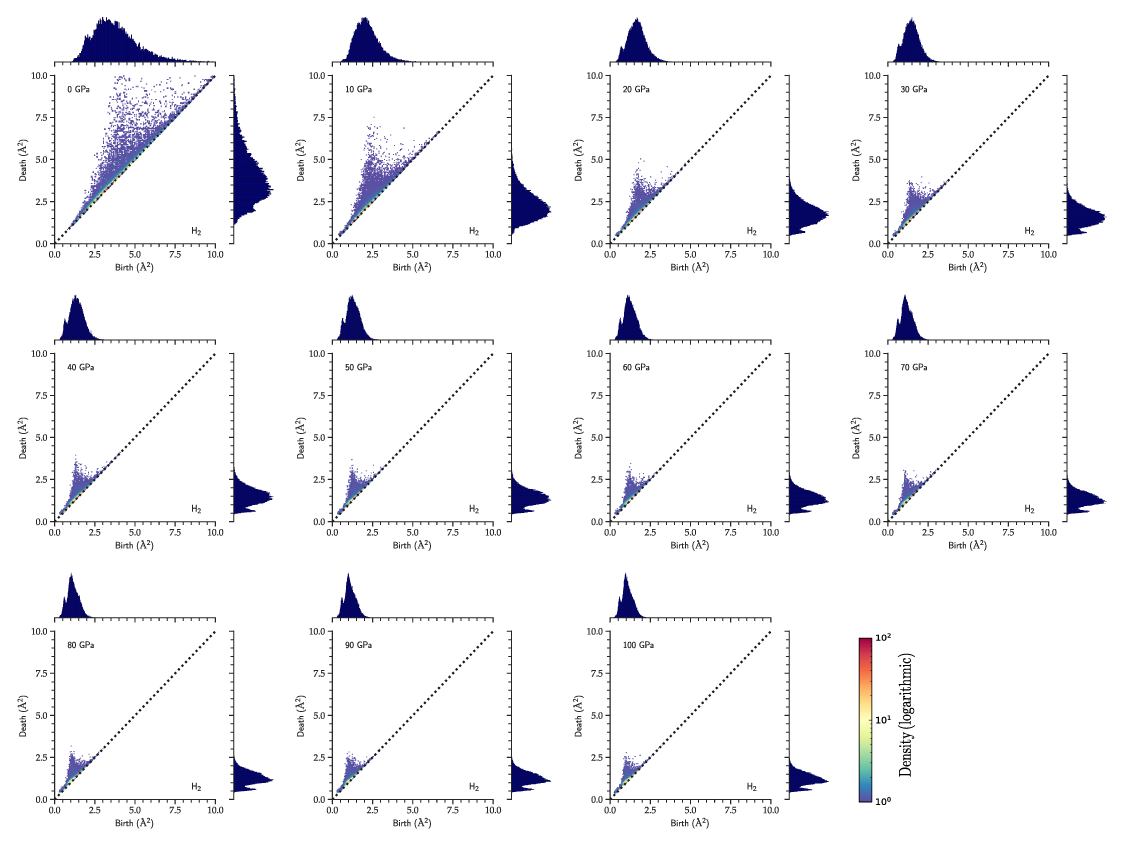}
\caption{Two-dimensional persistence diagram of silica glass at different selected pressures along the compression trajectory for $P = 0$~GPa to $P = 100$~GPa at an increment of 10~GPa. The radii used for the Si/O atoms are given in Tab.~\ref{stab:tab1}.}
\label{figs:pd2_r}
\end{figure*}

\begin{figure*}[htb!]
\centering
\includegraphics[width=\textwidth]{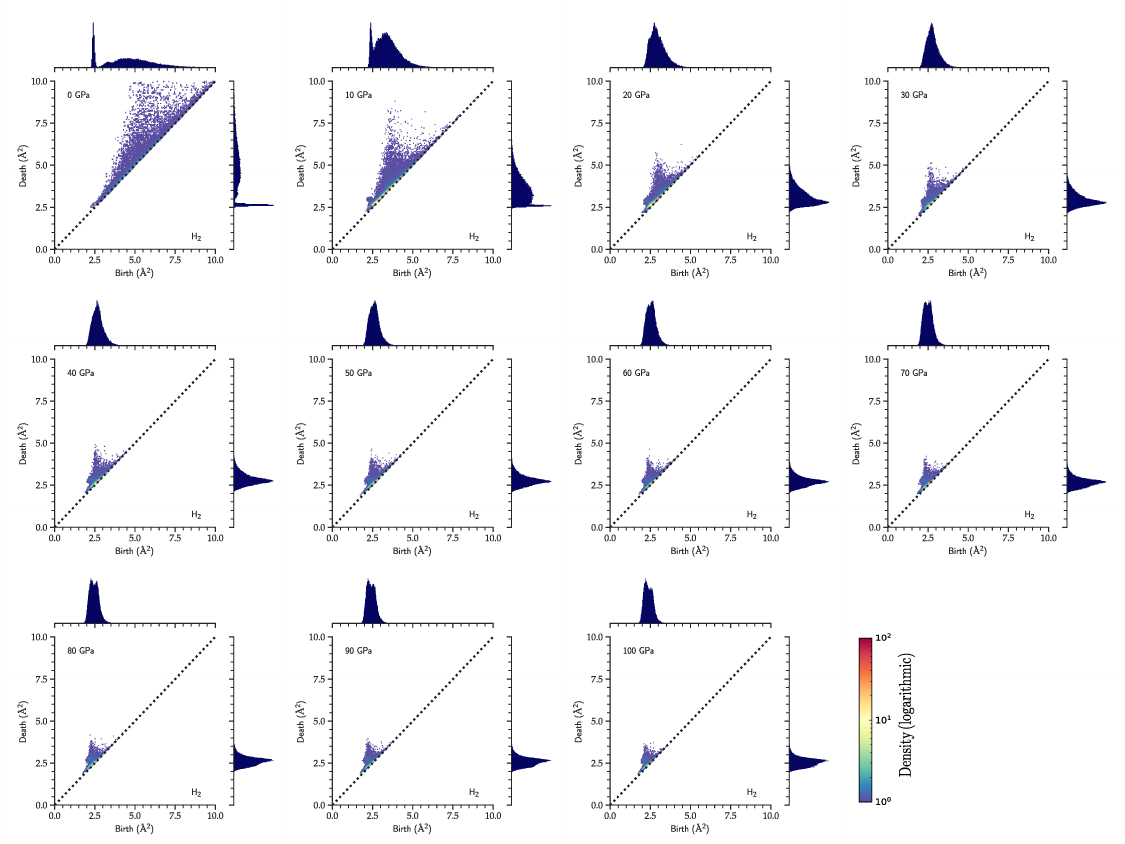}
\caption{O-centric two-dimensional persistence diagram of silica glass at different selected pressures along the compression trajectory for $P = 0$~GPa to $P = 100$~GPa at an increment of 10~GPa. The radii used for the O atoms are $r = 0$~\text{\AA}.}
\label{figs:pd2_O_0}
\end{figure*}

\begin{figure*}[htb!]
\centering
\includegraphics[width=\textwidth]{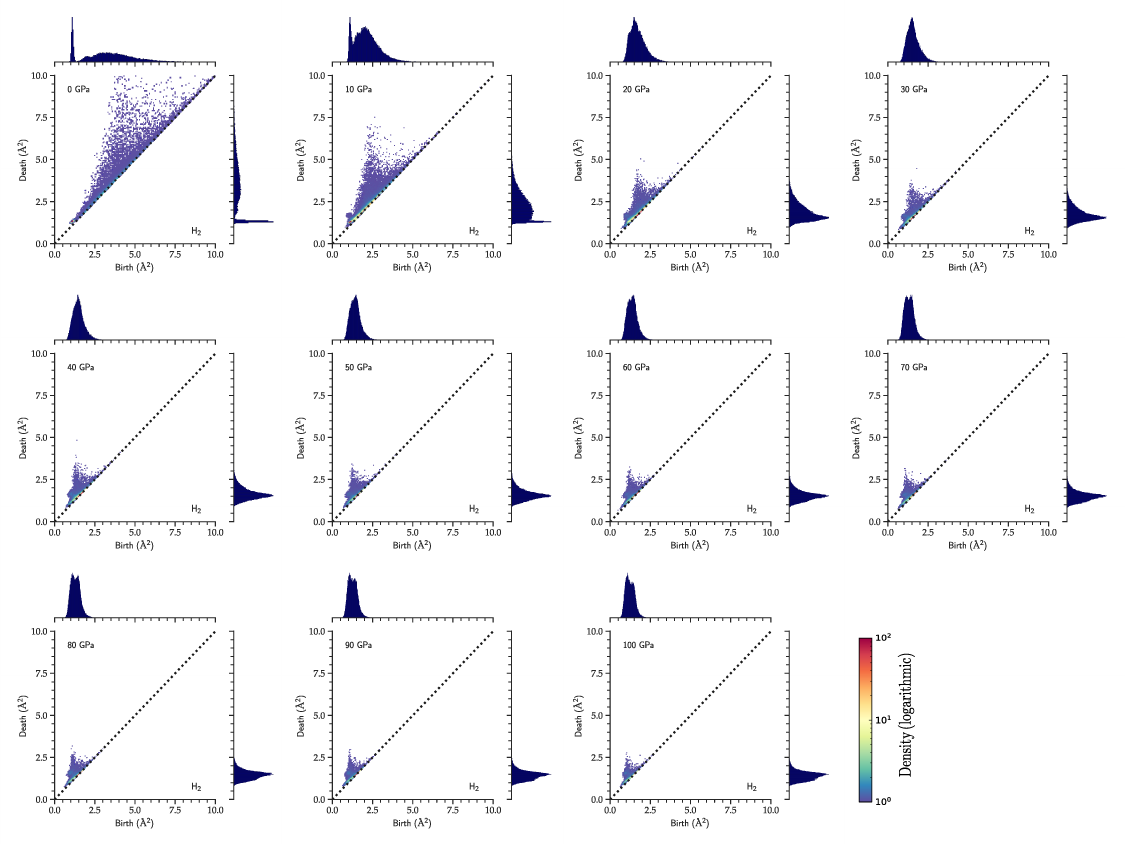}
\caption{O-centric two-dimensional persistence diagram of silica glass at different selected pressures along the compression trajectory for $P = 0$~GPa to $P = 100$~GPa at an increment of 10~GPa. The radii used for the O atoms are given in Tab.~\ref{stab:tab1}.}
\label{figs:pd2_O_r}
\end{figure*}

\begin{figure*}[htb!]
\centering
\includegraphics[width=\textwidth]{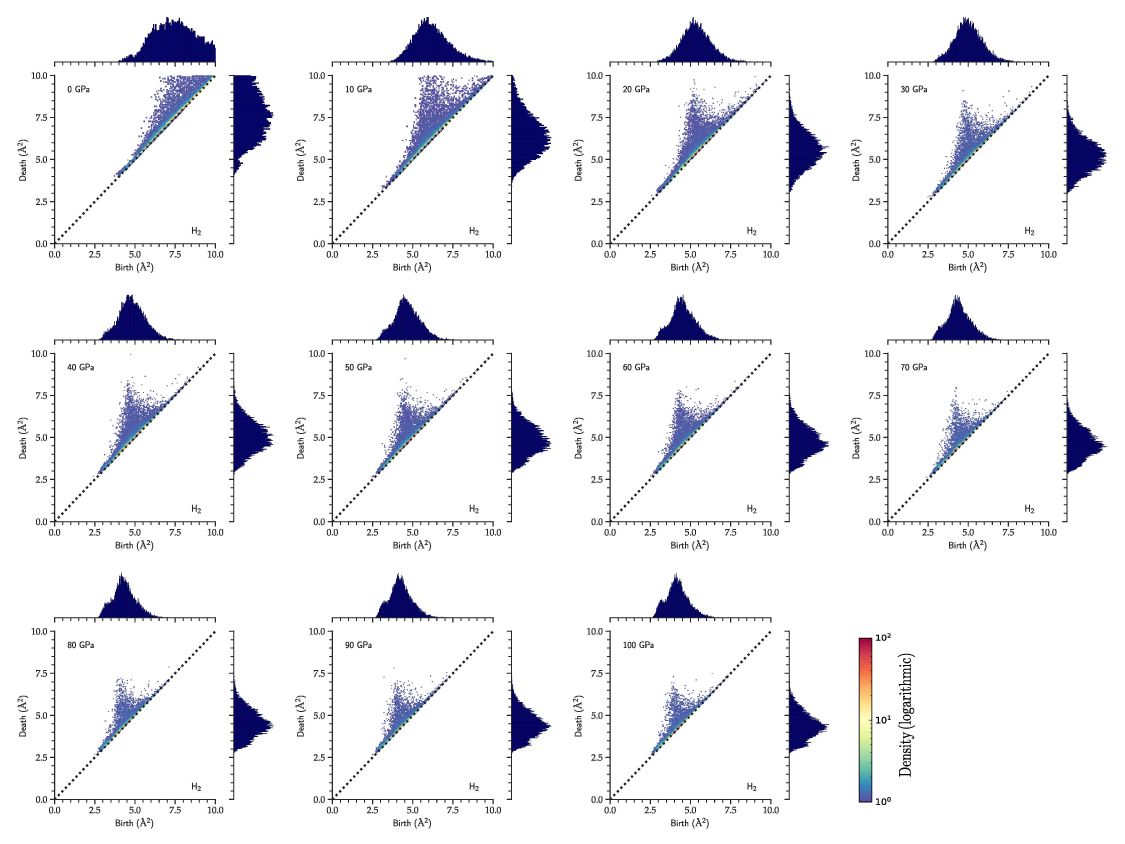}
\caption{Si-centric two-dimensional persistence diagram of silica glass at different selected pressures along the compression trajectory for $P = 0$~GPa to $P = 100$~GPa at an increment of 10~GPa. The radii used for the Si atoms are $r = 0$~\text{\AA}.}
\label{fig:pd2_Si_r0}
\end{figure*}

\begin{figure*}[htb!]
\centering
\includegraphics[width=\textwidth]{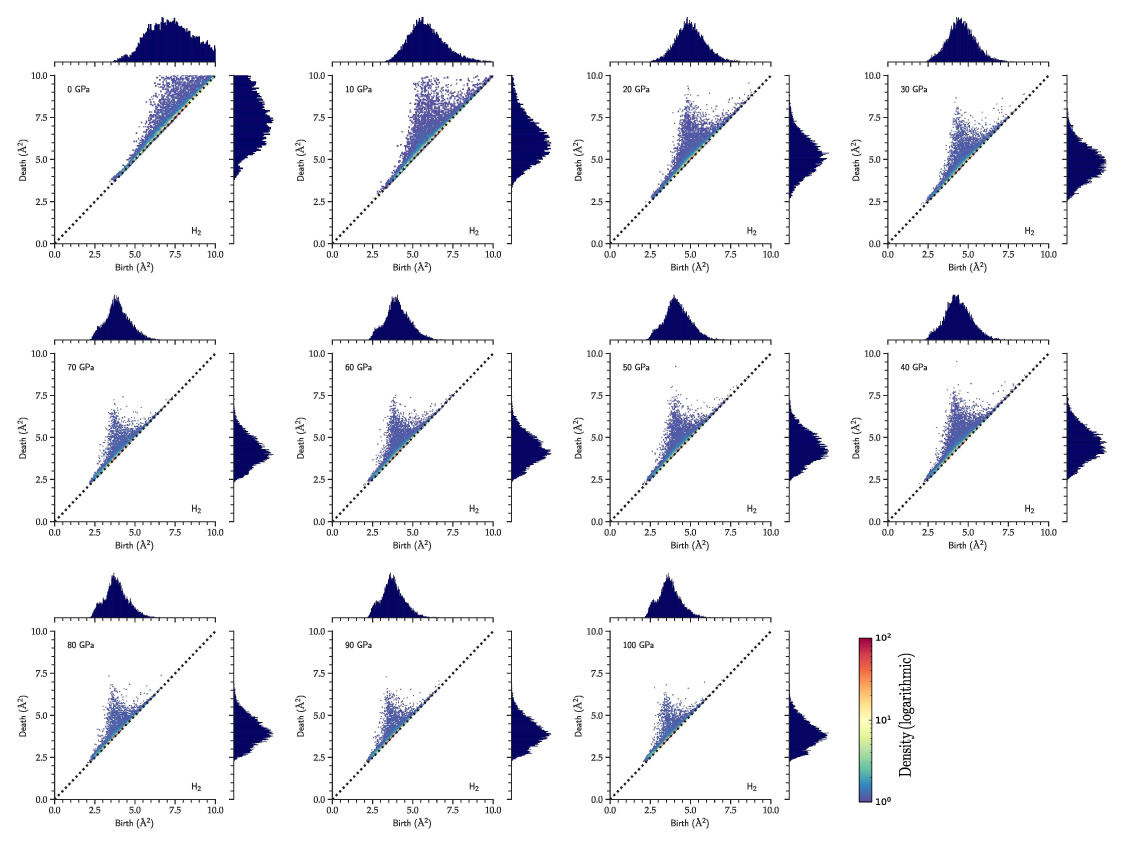}
\caption{Si-centric two-dimensional persistence diagram of silica glass at different selected pressures along the compression trajectory for $P = 0$~GPa to $P = 100$~GPa at an increment of 10~GPa. The radii used for the Si atoms are given in Tab.~\ref{stab:tab1}.}
\label{figs:pd2_Si_r}
\end{figure*}

\end{document}